\documentclass[aps,prd,reprint,superscriptaddress,nofootinbib]{revtex4-2}

\usepackage{amsmath}
\usepackage{amssymb}
\usepackage{bm}
\usepackage{graphicx}
\usepackage{booktabs}
\usepackage{hyperref}
\usepackage{slashed}

\newcommand{\GeV}{\,\mathrm{GeV}}

\newcommand{\pt}{p_{T}}
\newcommand{\ttbar}{\ensuremath{t\bar{t}}}
\newcommand{\ttbb}{\ensuremath{t\bar{t}b\bar{b}}}
\newcommand{\ww}{\ensuremath{WW}}
\newcommand{\zz}{\ensuremath{ZZ}}
\newcommand{\wwz}{\ensuremath{WWZ}}
\newcommand{\zzz}{\ensuremath{ZZZ}}

\begin{document}
		\title{Search for Heavy Charged Higgs Bosons in Boosted Top and Bottom Quark Final States at a 1.4 TeV Future Lepton Collider
		}
	
	\author{Majid Hashemi}
	\affiliation{Department of Physics, College of Science, Shiraz University, Shiraz, 71946-84795, Iran}
	
	\date{\today}
	
\begin{abstract}
The discovery potential for heavy charged Higgs bosons $H^\pm$ decaying into a top-bottom quark pair is investigated at a future $e^+e^-$ collider operating at $\sqrt{s}=1.4~\mathrm{TeV}$. The study considers charged Higgs masses between 200 and 600 GeV within all four Yukawa types of the two-Higgs-doublet model (2HDM). The boosted $tb$ final state is explored using complementary cut-based and machine-learning analyses based on boosted decision trees (BDT). The expected signal significance and sensitivity contours in the $(m_{H^\pm},\tan\beta)$ plane are obtained for each Yukawa type. The resulting sensitivity extends to regions of the 2HDM parameter space beyond the projected reach of the High-Luminosity LHC (HL-LHC), demonstrating the complementary potential of a future high-energy lepton collider to probe heavy charged Higgs bosons.
\end{abstract}

	\maketitle
	
	\section{Introduction}
	\label{sec:intro}
	
	The discovery of a Higgs boson with a mass of about $125~\GeV$ at the Large Hadron Collider (LHC) by the ATLAS and CMS collaborations \cite{ATLAS:2012yve, CMS:2012qbp} completed the particle content of the Standard Model (SM) and confirmed the mechanism of electroweak symmetry breaking (EWSB) proposed half a century earlier \cite{Higgs1, Higgs2, Higgs3, Englert1, Kibble1, Kibble2}. Since then, the central question of particle physics has shifted from discovery to characterisation: is the observed scalar the only one, or is it part of an extended Higgs sector? Precision measurements of its couplings \cite{LHC1, LHC6, LHC9}, its spin and CP properties \cite{LHC2, LHC3, LHC4, LHC7}, and its self-interaction are all consistent with the SM predictions within the current experimental accuracy, but the SM itself leaves several fundamental questions unanswered. It provides no dark matter candidate, no explanation for the matter--antimatter asymmetry of the Universe, no mechanism for neutrino masses, and no resolution of the hierarchy problem. Moreover, the Higgs sector of the SM is introduced in an ad hoc manner: a single complex doublet is the minimal choice, but nothing in the gauge structure of the theory requires it.
	
	Among the possible extensions of the scalar sector, the Two-Higgs-Doublet Model (2HDM) \cite{2hdm1, Branco:2011iw, hhg, 2hdm_TheoryPheno} is one of the simplest and most compelling. It enlarges the SM Higgs sector by a second complex $SU(2)_L$ doublet, yielding five physical states after EWSB: two CP-even scalars $h$ and $H$, one CP-odd pseudoscalar \(A\), and a pair of charged scalars \(H^\pm\). The model is motivated by a variety of theoretical considerations. It arises naturally in supersymmetric extensions of the SM \cite{MSSM1, MSSM2, MSSM4}, where the holomorphy of the superpotential requires at least two doublets. It provides a viable dark matter candidate in the inert-doublet variant \cite{IDM0, IDM1, IDM3}. It can accommodate a strongly first-order electroweak phase transition \cite{TypeI_LHC}, which is a prerequisite for electroweak baryogenesis. And, depending on the Yukawa structure, it can explain the observed fermion mass hierarchies and the pattern of flavour mixing. The 2HDM is therefore both a benchmark for Higgs-sector extensions and a realistic framework in which to interpret experimental data.
	
	The most general 2HDM scalar potential contains flavour-changing neutral currents (FCNCs) at tree level, which are strongly constrained by experiment. The standard remedy \cite{2hdm2, 2hdm3} is to impose a discrete $Z_2$ symmetry that allows each fermion type to couple to only one of the two doublets. This requirement defines four natural variants of the model, denoted Type-I, Type-II, Lepton-Specific, and Flipped. The phenomenology of the charged Higgs boson depends strongly on this choice, because the $H^\pm$ Yukawa couplings are proportional to the fermion masses multiplied by either $\tan\beta$ or $\cot\beta$, where $\tan\beta$ is the ratio of the vacuum expectation values of the two doublets. In the Type-I model, all Yukawa couplings are proportional to $\cot\beta$, so that for $\tan\beta \gtrsim 1$ the charged Higgs couples most strongly to the top and bottom quarks. In the Type-II and Flipped models, the down-type quark couplings are proportional to $\tan\beta$, leading to a strong enhancement of the \(H^\pm \to tb\) and $H^\pm \to \tau\nu$ widths at large $\tan\beta$. The Lepton-Specific model instead enhances the $H^\pm \to \tau\nu$ channel while leaving the quark couplings $\cot\beta$-suppressed. In this paper the discussion is kept general and results are presented for the four Yukawa types, comparing the reach of a future lepton collider with the current LHC exclusions.
	
	The charged Higgs boson has been the subject of intense experimental searches. At the LHC, the dominant production mechanisms are the associated production with a top quark, $bg \to tH^-$, and the decay of a top quark into a charged Higgs and a bottom quark, $t \to H^+ b$. These have been searched for extensively in the $\tau\nu$, $cs$, and \(tb\) decay channels by both ATLAS \cite{ATLAS:chnew1, ATLAS:chnew2, ATLAS:chnew5, ATLAS:2024CH} and CMS \cite{CMS:chnew1, CMS:chnew4, CMS:chnew6}. For $m_{H^\pm} \gtrsim m_t$, the decay \(H^\pm \to tb\) becomes kinematically accessible and quickly dominates in the Type-II and Flipped models, because the $H^\pm t b$ coupling grows with $\tan\beta$. The resulting final state, however, is buried under an enormous \(t\bar{t}\) background at a hadron collider, and the current LHC limits on the $tb$ channel remain relatively weak in these types. The most stringent limits at present come from the $\tau\nu$ channel in the Type-II and Lepton-Specific models, which is statistically cleaner but only sensitive to a limited region of parameter space. A definitive test of the 2HDM across the full parameter space therefore requires a different experimental environment.
	
	Future lepton colliders offer such an environment. The Compact Linear Collider (CLIC) \cite{cliccdr, CLICdp, CLICdet}, the International Linear Collider \cite{ILC, ilchiggs, ILCEnergy}, the Future Circular Collider in its $e^+e^-$ mode \cite{FCC-ee1, FCC-ee2, FCC-hh}, and the Circular Electron Positron Collider \cite{CEPC1, CEPC2} are all designed to collide electrons and positrons at centre-of-mass energies from the $Z$ pole up to several TeV. At these machines, the initial state is electroweak and colour-singlet, so that the QCD background that dominates at the LHC is absent. The clean environment, the possibility of polarised beams, and the precise knowledge of the initial state all combine to give lepton colliders a decisive advantage for many searches for new physics, particularly for processes that are kinematically suppressed or swamped by QCD at hadron colliders. The charged Higgs pair-production process $e^+e^- \to H^+H^-$, in particular, has a large and model-independent cross section that depends only on the gauge couplings and the charged Higgs mass. This process is therefore an ideal probe of the charged Higgs sector across all Yukawa types, and it is the one studied in this paper.
	
	The final state of $e^+e^- \to H^+H^-$ depends on the decay mode of the charged Higgs. For $m_{H^\pm} \gtrsim m_t$, the dominant decay is \(H^\pm \to tb\) in most of the parameter space, which gives a four-body final state $t\bar{t}b\bar{b}$. This state is characterised by a striking experimental signature: four \(b\)-jets, four light jets from the hadronic \(W\) decays, and two boosted top quarks that can be reconstructed as large-radius jets. The presence of four \(b\)-jets is the key feature that distinguishes the signal from the dominant \(t\bar{t}\) background, which contains only two \(b\)-jets. The \(t\bar{t}b\bar{b}\) background, where the two additional \(b\)-jets arise from gluon splitting, is also present, but its cross section is small at lepton colliders. The signal is therefore characterised by a sharp resonance peak in the invariant mass of the $tb$ system, which can be reconstructed from the four-jet and large-radius-jet information, combined with a distinctive four-\(b\)-jet multiplicity.
	
	In this paper a detailed feasibility study is presented for the discovery of a heavy charged Higgs boson in the \(H^\pm \to tb\) channel at a $\sqrt{s} = 1400~\GeV$ lepton collider, using the CLIC detector concept as a benchmark. The signal and the main Standard Model backgrounds are simulated, a cut-based analysis is developed that exploits the four-\(b\)-jet signature and the boosted top-quark reconstruction, and a four-body kinematic correction is applied that enforces the invariant mass of the $b\bar{b}t\bar{t}$ system to equal the centre-of-mass energy. It is shown that this procedure yields a clear resonance peak in the reconstructed charged Higgs mass, and the expected statistical significance is estimated for nine benchmark mass points in the range $200 \le m_{H^\pm} \le 600~\GeV$. The performance of jet substructure observables, such as $\tau_{32}$ and the large-radius jet mass, in separating the signal from the \(t\bar{t}\) background is also explicitly studied, and it is shown that the discrimination is dominated by the extra \(b\)-jets rather than by top-tagging. Finally, the projected limits are compared with the current LHC exclusions in the four Yukawa types of the 2HDM. The paper is organised as follows. In Sec.~\ref{sec:theory} the relevant features of the 2HDM and the current LHC constraints are reviewed. In Sec.~\ref{sec:simulation} the event generation and detector simulation are described. In Sec.~\ref{sec:analysis} the analysis strategy is presented. In Sec.~\ref{sec:top-tagging} the top-tagging performance and the object multiplicities are studied. In Sec.~\ref{sec:results} the results are presented, and in Sec.~\ref{sec:conclusion} the conclusions are summarised.
	
	\section{Theoretical Framework}
	\label{sec:theory}
	
	\subsection{The 2HDM scalar sector}
	
	The 2HDM extends the SM by a second complex $SU(2)_L$ scalar doublet. The two doublets are conventionally written as
	\begin{equation}
		\Phi_i = \begin{pmatrix} \phi_i^+ \\ \frac{1}{\sqrt{2}}(v_i + \phi_i^0 + i\chi_i) \end{pmatrix}, \qquad i = 1, 2,
	\end{equation}
	where $v_1$ and $v_2$ are the vacuum expectation values of the two doublets, satisfying $v_1^2 + v_2^2 = v^2 \approx (246~\GeV)^2$, and $\tan\beta = v_2/v_1$ is the ratio of their vacuum expectation values. The most general renormalisable scalar potential consistent with gauge invariance and the softly broken $Z_2$ symmetry is
	\begin{equation}
		\begin{aligned}
			V(\Phi_1, \Phi_2) &= m_{11}^2 \Phi_1^\dagger \Phi_1 + m_{22}^2 \Phi_2^\dagger \Phi_2 - m_{12}^2 (\Phi_1^\dagger \Phi_2 + \text{h.c.}) \\
			&\quad + \frac{\lambda_1}{2} (\Phi_1^\dagger \Phi_1)^2 + \frac{\lambda_2}{2} (\Phi_2^\dagger \Phi_2)^2 + \lambda_3 (\Phi_1^\dagger \Phi_1)(\Phi_2^\dagger \Phi_2) \\
			&\quad + \lambda_4 (\Phi_1^\dagger \Phi_2)(\Phi_2^\dagger \Phi_1) + \frac{\lambda_5}{2} \left[ (\Phi_1^\dagger \Phi_2)^2 + \text{h.c.} \right],
		\end{aligned}
		\label{eq:potential}
	\end{equation}
	where $m_{12}^2$ is the soft-breaking parameter of the $Z_2$ symmetry and the $\lambda_i$ are real quartic couplings in the CP-conserving limit. After EWSB, the scalar spectrum consists of two CP-even neutral states $h$ and $H$, one CP-odd neutral state \(A\), and a pair of charged states \(H^\pm\). The physical states are obtained by rotating the interaction eigenstates through the angles $\alpha$ and $\beta$, where $\alpha$ diagonalises the CP-even mass matrix. The masses of the physical states are
	\begin{align}
		m_h^2 &= \lambda_1 v_1^2 + \lambda_2 v_2^2 - \sqrt{(\lambda_1 v_1^2 - \lambda_2 v_2^2)^2 + 4 \lambda_{345}^2 v_1^2 v_2^2}, \\
		m_H^2 &= \lambda_1 v_1^2 + \lambda_2 v_2^2 + \sqrt{(\lambda_1 v_1^2 - \lambda_2 v_2^2)^2 + 4 \lambda_{345}^2 v_1^2 v_2^2}, \\
		m_A^2 &= \frac{m_{12}^2}{v_1 v_2} v^2 - \lambda_5 v^2, \\
		m_{H^\pm}^2 &= \frac{m_{12}^2}{v_1 v_2} v^2 - \frac{1}{2}(\lambda_4 + \lambda_5) v^2,
	\end{align}
	with $\lambda_{345} = \lambda_3 + \lambda_4 + \lambda_5$. The light CP-even state $h$ is identified with the observed 125~GeV Higgs boson, and the remaining states $(H, A, H^\pm)$ are the heavy or new scalars of the model. The parameters $\alpha$ and $\beta$ are related by $\cos(\beta-\alpha)$, which controls the alignment of the light Higgs couplings with their SM values: in the alignment limit $\cos(\beta-\alpha) \to 0$, the light Higgs $h$ couples exactly as in the SM, and the heavy states decouple from the gauge sector. In this paper the alignment limit is assumed.
	
	\subsection{Yukawa structure and the four 2HDM types}
	
	The most general Yukawa interactions of the two doublets with the SM fermions contain tree-level FCNCs, which are strongly constrained by the observed $K^0$--$\bar{K}^0$ and $B^0$--$\bar{B}^0$ mixing. The standard remedy is to impose a discrete $Z_2$ symmetry that allows each fermion type to couple to only one of the two doublets. The four natural variants of the model are:
	\begin{itemize}
		\item \textbf{Type-I:} all quarks and charged leptons couple to $\Phi_2$.
		\item \textbf{Type-II:} up-type quarks couple to $\Phi_2$, down-type quarks and charged leptons to $\Phi_1$.
		\item \textbf{Lepton-Specific:} quarks couple to $\Phi_2$, leptons to $\Phi_1$.
		\item \textbf{Flipped:} up-type quarks and leptons couple to $\Phi_2$, down-type quarks to $\Phi_1$.
	\end{itemize}
	The charged Higgs interaction with the third-generation quarks can be written in a unified way as
	\begin{equation}
		\mathcal{L}_{H^\pm} \supset \frac{g}{\sqrt{2}m_W} H^+ \bar{t} \left( \xi_t m_t P_L + \xi_b m_b P_R \right) b + \text{h.c.},
		\label{eq:yukawa}
	\end{equation}
	where the coefficients $\xi_t$ and $\xi_b$ depend on the Yukawa type and on $\tan\beta$:
	\begin{align}
		\text{Type-I:} \quad &\xi_t = \cot\beta, \quad &&\xi_b = \cot\beta, \\
		\text{Type-II:} \quad &\xi_t = \cot\beta, \quad &&\xi_b = \tan\beta, \\
		\text{Lepton-Specific:} \quad &\xi_t = \cot\beta, \quad &&\xi_b = \cot\beta, \\
		\text{Flipped:} \quad &\xi_t = \cot\beta, \quad &&\xi_b = \tan\beta.
	\end{align}
	For the decay \(H^\pm \to tb\), the width is
\begin{equation}
	\begin{aligned}
		\Gamma(H^\pm\to tb)
		={}&
		\frac{N_c g^2 |V_{tb}|^2m_{H^\pm}}
		{32\pi m_W^2}
		\lambda^{1/2}(1,r_t,r_b)
		\\
		&\times
		\Bigg[
		(\xi_t^2m_t^2+\xi_b^2m_b^2)
		(1-r_t-r_b)
		\\
		&\qquad
		-\frac{4\xi_t\xi_bm_t^2m_b^2}
		{m_{H^\pm}^2}
		\Bigg],
	\end{aligned}
	\label{eq:charged_higgs_tb_width}
\end{equation}
where \(N_c=3\) is the number of QCD colours, \(g\) is the \(SU(2)_L\) gauge coupling, \(V_{tb}\) is the corresponding Cabibbo--Kobayashi--Maskawa matrix element. The dimensionless mass parameters are defined as
\begin{equation}
	r_t=\frac{m_t^2}{m_{H^\pm}^2},
	\qquad
	r_b=\frac{m_b^2}{m_{H^\pm}^2},
\end{equation}
and the kinematic function is
\begin{equation}
	\lambda(x,y,z)=x^2+y^2+z^2-2xy-2xz-2yz.
\end{equation}
The coefficients \(\xi_t\) and \(\xi_b\) parameterise the charged-Higgs couplings to the top and bottom quarks, respectively, and depend on the Yukawa type of the two-Higgs-doublet model and on \(\tan\beta\), as specified in Eqs.~(8)--(11).

In the Type-I model, all fermionic couplings of the charged Higgs boson are proportional to \(\cot\beta\). Consequently, both the \(tb\) and \(\tau\nu\) partial widths are suppressed at large \(\tan\beta\), while their relative importance is not directly determined by the \(\tan\beta\) dependence of the Yukawa couplings. In the Type-II and Flipped models, the coupling \(\xi_b=\tan\beta\) enhances the \(tb\) width at large \(\tan\beta\). In contrast, the \(tb\) width remains \(\cot\beta\)-suppressed in the Lepton-Specific model. The \(\tau\nu\) channel is governed by \(\xi_\tau=\cot\beta\) in the Type-I and Flipped models, and by \(\xi_\tau=\tan\beta\) in the Type-II and Lepton-Specific models. Therefore, the \(\tau\nu\) channel is enhanced at large \(\tan\beta\) in the Type-II and Lepton-Specific models, whereas it is suppressed in the Type-I and Flipped models. These distinctions are central to the phenomenology of the charged Higgs at colliders, and they motivate the model-independent approach adopted in this paper.
	
	\subsection{Production at lepton colliders}
	
	At an $e^+e^-$ collider, the charged Higgs bosons are pair-produced through the $s$-channel exchange of a photon or a $Z$ boson:
	\begin{equation}
		e^+e^- \to \gamma^*/Z^* \to H^+H^-.
	\end{equation}
	The cross section depends only on the electroweak gauge couplings and on the charged Higgs mass, and is independent of $\tan\beta$ and of the Yukawa type. This is a decisive advantage over the LHC, where the production mechanisms (associated production with a top quark, top decay, and vector-boson fusion) all depend on the Yukawa couplings and therefore on the details of the 2HDM parameter space. The cross section for $e^+e^- \to H^+H^-$ at $\sqrt{s} = 1400~\GeV$ is of order a few femtobarns for charged Higgs masses in the range $300$--$600~\GeV$, which yields a large number of events at the integrated luminosities foreseen for CLIC. The clean initial state, the absence of QCD radiation, and the precise knowledge of the centre-of-mass energy all combine to make the charged Higgs pair-production process an ideal benchmark for the 2HDM at future lepton colliders, and they allow a direct comparison of the reach across the four Yukawa types.
	
	\subsection{Current LHC constraints on the charged Higgs}
	
The charged Higgs boson has been searched for extensively at the LHC. The dominant production mechanisms are the associated production with a top quark, \(bg \to tH^-\), and the top-quark decay \(t \to H^+ b\). The decay channels that have been exploited include \(\tau\nu\), \(cs\), and \(tb\).

In all four Yukawa types, the LHC sensitivity is affected by the substantial backgrounds associated with hadronic collisions and by the dependence of the charged-Higgs production rates on the Yukawa couplings. A future lepton collider, by contrast, can exploit the pair-production process \(e^+e^- \to H^+H^-\), whose tree-level production cross section is determined by the electroweak gauge interactions and is independent of the Yukawa type and \(\tan\beta\). Moreover, the clean experimental environment of a lepton collider offers significant advantages in controlling backgrounds, particularly those associated with QCD interactions.

\begin{figure}[t]
	\centering
	\includegraphics[width=0.95\linewidth]{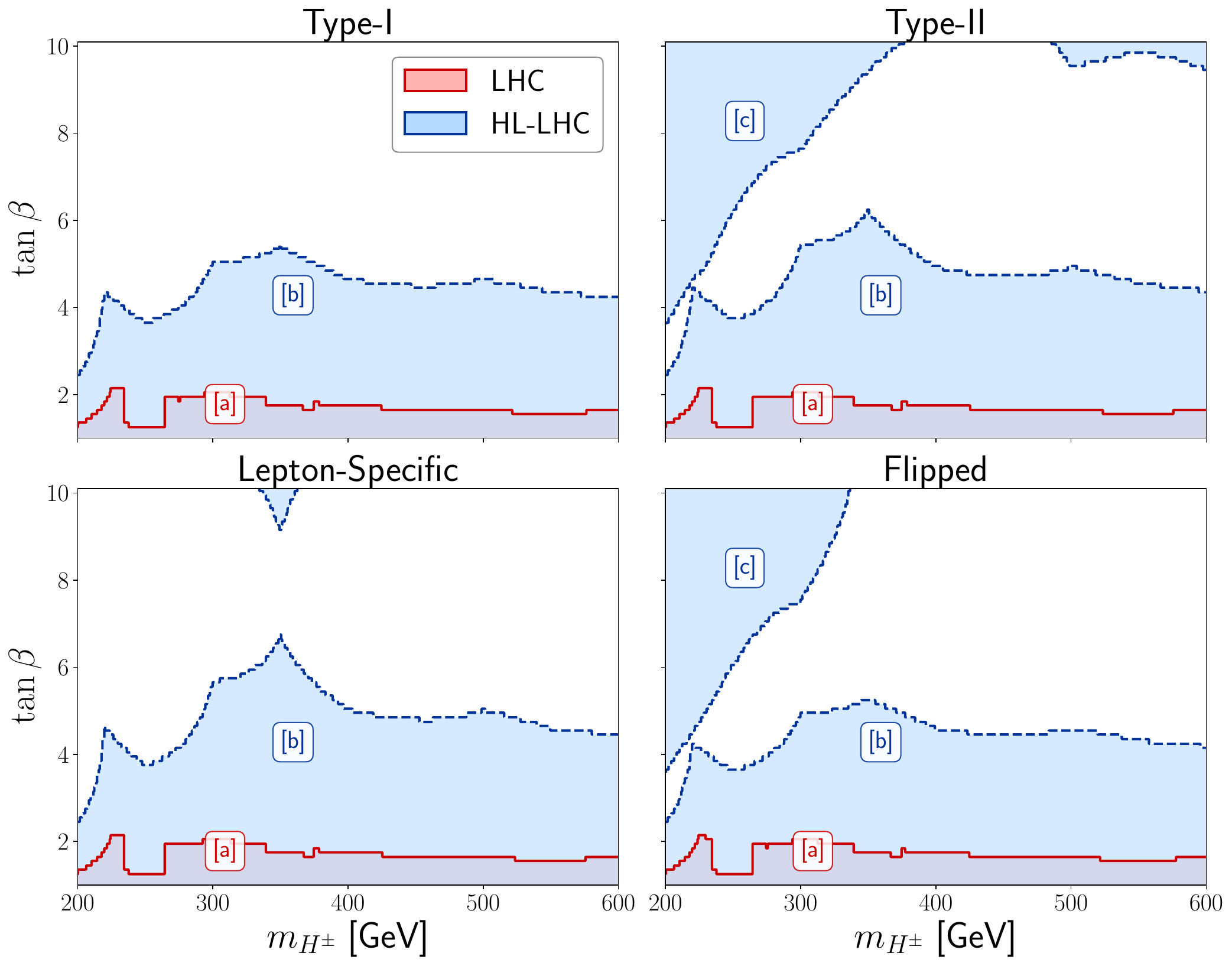}
\caption{HiggsTools exclusion contours in the
	\((m_{H^\pm}, \tan\beta)\) plane for the four 2HDM types.
	The red shaded regions (solid contours) are excluded by
	the current LHC data, while the blue shaded regions
	(dashed contours) correspond to the HL-LHC projections.
	The labels [a], [b], and [c] indicate the experimental
	results used to obtain the corresponding exclusion
	contours: ATLAS \(H^\pm\to tb\) search~\cite{ATLAS:chnew2},
	CMS \(H^\pm\to tb\) search~\cite{CMS:chnew1}, and ATLAS
	\(H^\pm\to\tau\nu\) search~\cite{ATLAS:chnew1},
	respectively.}
	\label{fig:higgstools}
\end{figure}

The current exclusion contours in the \((m_{H^\pm}, \tan\beta)\) plane, obtained using the \textsc{HiggsTools} package \cite{higgstools}, are shown in Fig.~\ref{fig:higgstools} for the LHC (red) and HL-LHC (blue, dashed). The current LHC exclusions are remarkably similar across the four Yukawa types, reaching approximately \(\tan\beta\lesssim 1.5\)--\(2\) over much of the displayed charged-Higgs mass range. The HL-LHC projections significantly extend the excluded regions, although their impact depends strongly on the Yukawa structure. In Type-I, the excluded region extends towards larger values of \(\tan\beta\), reaching approximately \(\tan\beta\sim 4\)--\(5\). A similar pattern is observed in the Lepton-Specific model, with a somewhat different contour shape. In Type-II and Flipped, the HL-LHC projections yield two distinct excluded regions, at low and high \(\tan\beta\), leaving an intermediate allowed region over much of the mass range considered. These results illustrate the importance of considering the different Yukawa structures when assessing the complementarity between hadron and future lepton colliders.
In the remainder of this paper, the reach of a future lepton collider is quantified and compared with the current LHC and HL-LHC exclusions shown in Fig.~\ref{fig:higgstools}.

	\section{Event Generation and Simulation}
	\label{sec:simulation}
	
The signal events \(e^+e^- \to H^+H^-\) are generated using \textsc{Whizard}~3.1 \cite{whizard1}, with the subsequent decays and parton showering handled by \textsc{Pythia}~8 \cite{pythia}. The detector response is simulated with \textsc{Delphes}~3 \cite{delphes1, overlay} using the CLICdet Stage~2 configuration \cite{CLICdet}, which implements the tracking and calorimeter resolutions of the CLICdet concept and includes the effect of beam-induced backgrounds. \\
Charged-particle tracks are reconstructed with an inverse transverse-momentum resolution of approximately \(1.3\times10^{-5}~\GeV^{-1}\) for central high-\(p_T\) tracks, as parametrised by the CLICdet momentum-smearing model. The electromagnetic and hadronic calorimeters are also simulated with the energy resolutions of the CLICdet performance studies. \\
At the Stage~2 energy, the effect of the \(\gamma\gamma \to \mathrm{hadrons}\) overlay background is modelled through the standard CLICdet jet-energy smearing, which applies a flat resolution of \(1\%\) in the barrel (\(|\eta| < 0.76\)) and \(5\%\) in the endcap (\(|\eta| \ge 0.76\)) to all jet collections. The same smearing is applied to the small-\(R\) jets and to the large-\(R\) top-jets, and the resulting jet energy resolution is at the few-percent level for the jet energies relevant to this analysis.\\
The nominal centre-of-mass energy of the collider is \(\sqrt{s}=1.4~\text{TeV}\). However, the actual collision energy varies from event to event due to the beam-energy spread and beamstrahlung, which lead to a reduction of the incoming electron and positron energies before the hard interaction. Consequently, the collisions take place over a luminosity spectrum extending from the nominal centre-of-mass energy towards lower energies. This effect is particularly relevant at high-energy linear colliders such as CLIC, where beamstrahlung produces a pronounced low-energy tail in the luminosity spectrum \cite{cliccdr}.

\begin{figure}[t]
	\centering
	\includegraphics[width=0.9\linewidth]{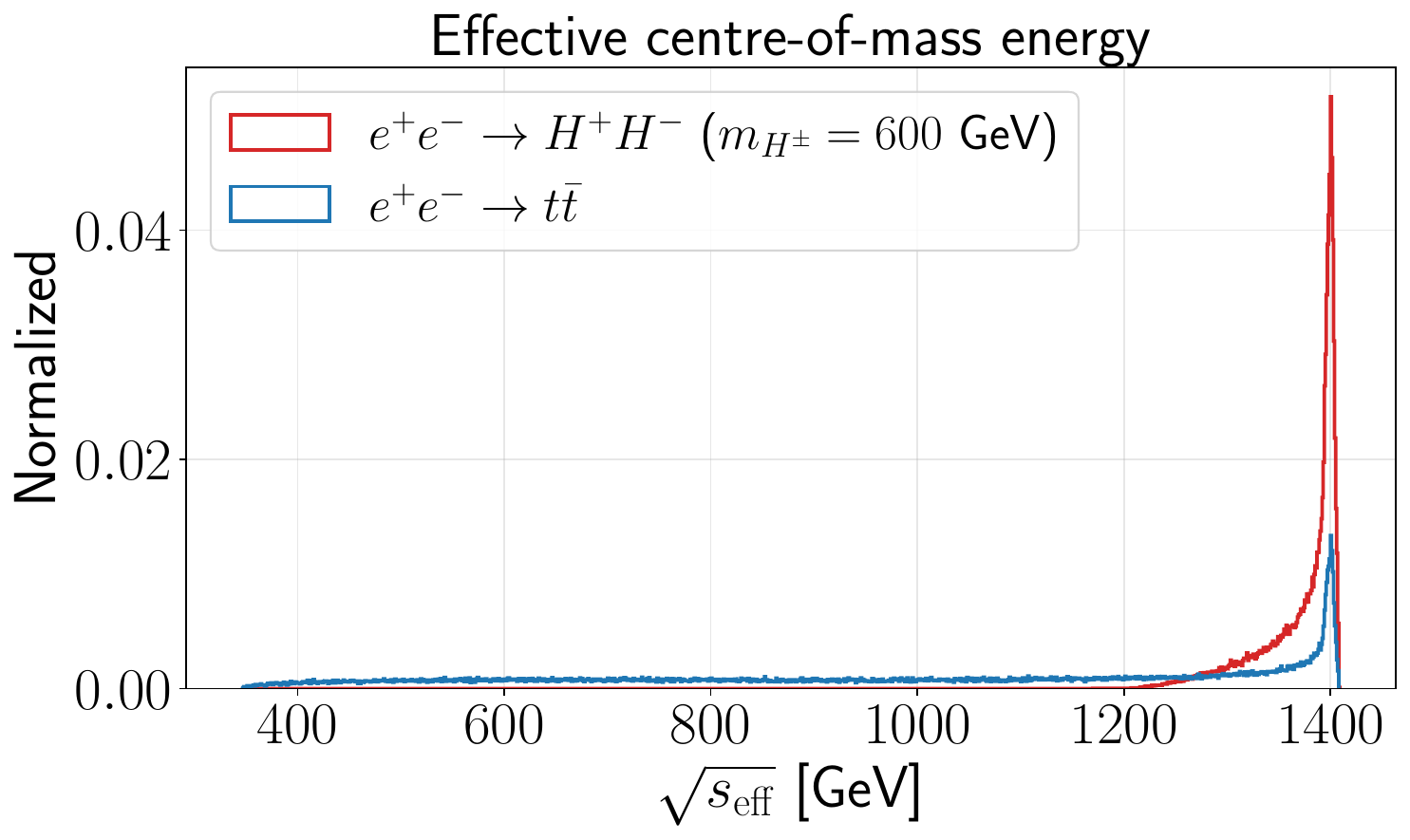}
	\caption{Effective centre-of-mass energy distributions, \(\sqrt{s_{\rm eff}}\), obtained from the incoming electron and positron four-momenta in the WHIZARD-generated LHE events for \(e^+e^-\to H^+H^-\) with \(m_{H^\pm}=600~\text{GeV}\) and \(e^+e^-\to t\bar{t}\). The lower kinematic boundaries correspond to the respective on-shell pair-production thresholds.}
	\label{fig:beamspectrum}
\end{figure}

The effective centre-of-mass energy, \(\sqrt{s_{\rm eff}}\), is evaluated for each event using the four-momenta of the incoming electron and positron recorded in the LHE files generated by WHIZARD:
\[
s_{\rm eff} = (p_{e^-}+p_{e^+})^2.
\]
The resulting distributions therefore reflect the event-by-event collision energy after the beam-energy spectrum has been taken into account. In particular, the distributions exhibit clear kinematic thresholds associated with the production of on-shell particle pairs. For the signal process, \(e^+e^-\to H^+H^-\), the minimum effective centre-of-mass energy is approximately \(2m_{H^\pm}\), while for the dominant \(t\bar{t}\) background it is approximately \(2m_t\). These thresholds are evident in the effective centre-of-mass energy distributions, as illustrated in Fig.~\ref{fig:beamspectrum}. For example, the \(m_{H^\pm}=600~\text{GeV}\) signal sample exhibits a lower kinematic boundary around \(1.2~\text{TeV}\), whereas the \(t\bar{t}\) distribution starts near the top-quark pair-production threshold. These lower boundaries arise from the phase-space requirements of the corresponding hard-scattering processes and are not imposed by an additional analysis-level selection. Events below the respective pair-production thresholds cannot produce the corresponding on-shell final states.\\
Nine signal mass points are simulated, with \(m_{H^\pm}\) ranging from 200 to 600~GeV in steps of 50~GeV. The main SM backgrounds considered are \(e^+e^- \to t\bar{t}\), \(t\bar{t}b\bar{b}\), \(WW\), \(ZZ\), \(WWZ\), and \(ZZZ\). The analysis is performed on a Monte Carlo sample corresponding to an integrated luminosity of \(\mathcal{L}_{\text{int}} = 1000~\text{fb}^{-1}\).
	
	\subsection{Cross Sections and Signal-to-Background Ratio}
	\label{sec:cross_sections}
	
\begin{figure}[t]
		\centering
		\includegraphics[width=0.9\linewidth]{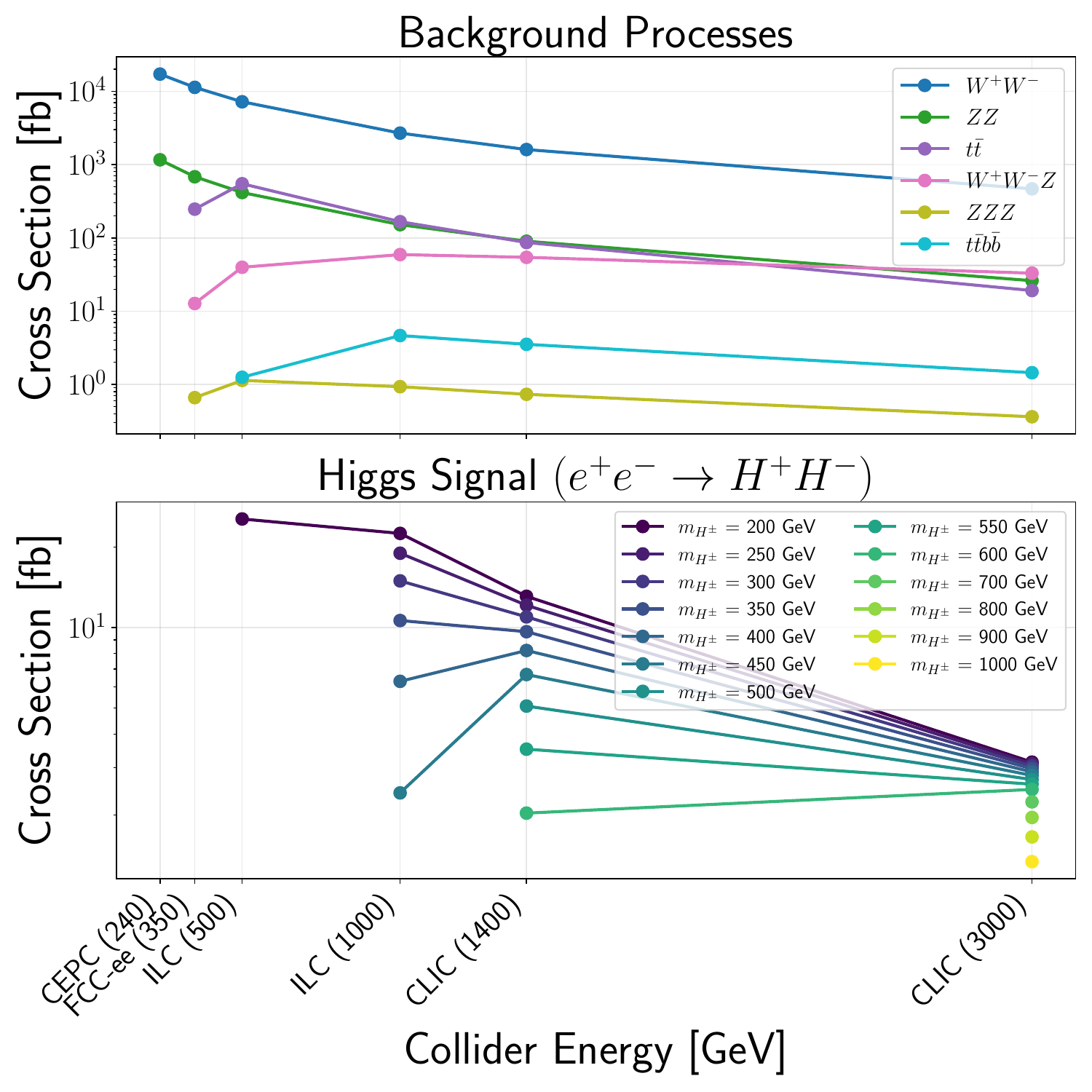}
		\caption{Cross sections for the dominant background processes (top) and for the Higgs signal (bottom) as a function of the collider centre-of-mass energy. The background is dominated by $\ww$, $\zz$ and $t\bar{t}$.}
		\label{fig:cross_sections}
	\end{figure}

	Before describing the event selection, it is instructive to examine the energy dependence of the signal and background cross sections and the resulting signal-to-background ratio. Fig.~\ref{fig:cross_sections} shows the cross sections for the dominant background processes (top) and for the Higgs signal (bottom) as a function of the collider center-of-mass energy. The background cross sections are dominated by the gauge boson pair production and $e^+e^- \to t\bar{t}$. The signal cross section for \(e^+e^- \to H^+H^-\) is determined by the electroweak gauge couplings and the available phase space. Consequently, mass points produced close to the kinematic threshold, \(m_{H^\pm}\simeq\sqrt{s}/2\), where the cross section is strongly suppressed by phase-space effects, can exhibit a significant increase in their production cross sections at higher centre-of-mass energies.
	
\begin{figure}[t]
		\centering
		\includegraphics[width=0.9\linewidth]{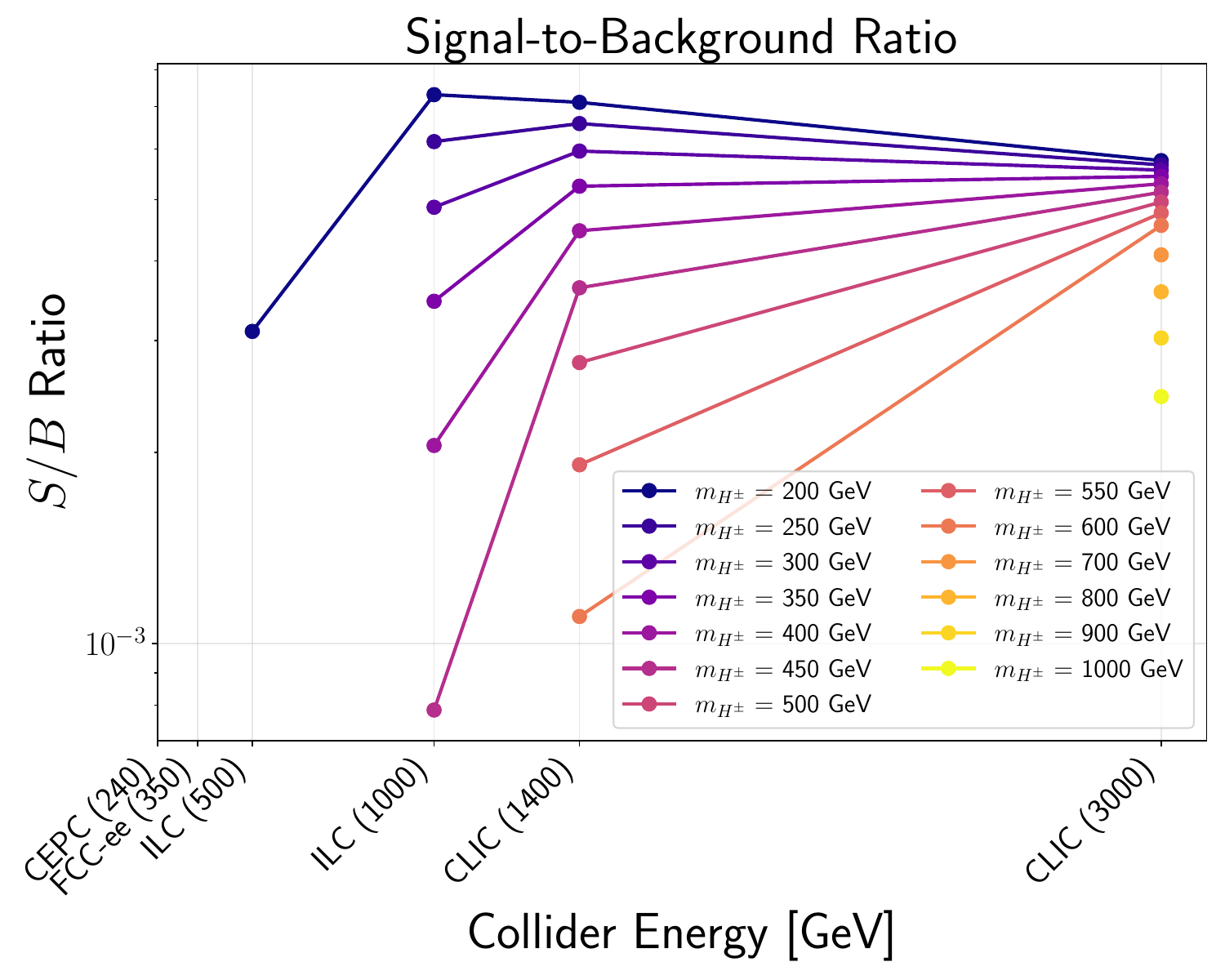}
		\caption{Signal-to-background ratio as a function of the collider centre-of-mass energy for representative charged Higgs masses. The ratio remains at the percent level even at the highest energies, motivating the aggressive cut-based selection of Sec.~\ref{sec:analysis}.}
		\label{fig:sb_ratio}
	\end{figure}

	The resulting signal-to-background ratio is shown in Fig.~\ref{fig:sb_ratio}. For light charged Higgs masses ($m_{H^\pm} \lesssim 300~\GeV$), the ratio is of order $10^{-3}$--$10^{-2}$, reflecting the small signal cross section and the large background. For heavier masses the ratio remains at the few-percent level even at the highest energies considered. This justifies the aggressive selection strategy of Sec.~\ref{sec:analysis}, which is designed to suppress the background by several orders of magnitude while retaining a substantial fraction of the signal.
	
	\section{Event Selection and Reconstruction}
	\label{sec:analysis}
	
	Jets are considered in the analysis if they satisfy \(\pt > 20~\GeV\) and \(|\eta| < 5\), while top-tagged jets are required to have \(\pt > 200~\GeV\) and \(|\eta| < 5\). The \(b\)-tagging efficiency of 90\% is chosen among the three 50\%, 70\%, and 90\% operating points.
	
	\subsection{Event Preselection}
	
	Jets are reconstructed using two different clustering algorithms. Large-radius top jets are clustered with the Cambridge–Aachen (C/A) algorithm \cite{CA} using a radius parameter of $R=2.0$, while the remaining jets are reconstructed with the Valencia algorithm \cite{valencia} using $R=0.5$. Jets passing the $b$-tagging requirement are classified as $b$-jets, while those failing it are classified as light jets. Events are retained if they contain at least two fat top jets, at least four $b$-jets, and at least four light jets. These requirements define the preselection stage and are applied sequentially, with the corresponding selection efficiencies recorded.
	
	\subsection{\(W\) Boson Reconstruction}
	
	Among the selected light jets, all distinct pairs of dijet combinations are considered. For each pair of dijets \((i,j)\) and \((k,l)\), the invariant masses \(m_{W1}\) and \(m_{W2}\) are computed, and the combination minimizing
	\begin{equation}
		\Delta m_{W} = |m_{W1} - m_{W}| + |m_{W2} - m_{W}|,
		\label{eq:deltamW}
	\end{equation}
	with \(m_{W} = 80~\GeV\), is chosen. The four-momenta of the two selected dijets are then rescaled so that their invariant masses equal \(m_{W}\) exactly. Events are accepted if both reconstructed \(W\) candidates satisfy \(|m_{W_i} - m_{W}| < 30~\GeV\).
	
	\subsection{Top Quark Reconstruction}
	
	The two \(W\) candidates are combined with the selected \(b\)-tagged jets. All distinct pairings of \(b\)-jets with the two \(W\) bosons are tested, and the combination minimizing
	\begin{equation}
		\Delta m_{\text{top}} = |m_{\text{top},1} - m_{\text{top}}|
		+ |m_{\text{top},2} - m_{\text{top}}|,
		\label{eq:deltamtop}
	\end{equation}
	with \(m_{\text{top}} = 173~\GeV\), is retained. The four-momenta of each \(W\)--\(b\) pair are rescaled so that the resulting top quark invariant masses equal \(m_{\text{top}}\). Events are accepted if both reconstructed top quarks satisfy \(|m_{\text{top},i} - m_{\text{top}}| < 40~\GeV\).
	
	\subsection{Charged Higgs Reconstruction and Kinematic Correction}
	
The remaining \(b\)-tagged jets, not used in the top quark reconstruction, are paired with the two reconstructed top quarks. A four-body kinematic correction is applied to the two \(b\)-jets and the two top quarks, enforcing the invariant mass of the \(b\bar{b}t\bar{t}\) system to equal \(\sqrt{s} = 1400~\GeV\). The two charged Higgs candidates are then formed as \(H^{\pm}_{1} = t_{1} + b_{2}\) and \(H^{\pm}_{2} = t_{2} + b_{1}\), after sorting the \(b\)-jets and top quarks by energy, and their invariant masses are recorded.

The kinematic correction enforces the constraint that the four-body system \(b_1 + b_2 + t_1 + t_2\) has an invariant mass equal to a fixed value \(\sqrt{s} = 1400~\GeV\). At this stage, the two reconstructed top quarks and the two remaining \(b\)-jets have already been identified. Their four-momenta are labelled as \(p_1, p_2, p_3, p_4\), where \(p_1, p_2\) are the two \(b\)-jets and \(p_3, p_4\) are the two top quarks, and each four-momentum is written in components \(p_i = (E_i, p_{ix}, p_{iy}, p_{iz})\). The correction assumes that each of the four objects is rescaled by a common, object-dependent factor \(c_i\) applied to all four components of its four-momentum, \(p_i \to c_i p_i\), which preserves the direction of each object in three-momentum space and scales its energy and momentum uniformly. Solving the resulting linear system gives the four correction factors as ratios of determinants,
\begin{equation}
	c_1 = \frac{s\, \mathcal{N}_1}{\mathcal{D}},~~
	c_2 = -\frac{s\, \mathcal{N}_2}{\mathcal{D}},~~
	c_3 = \frac{s\, \mathcal{N}_3}{\mathcal{D}},~~
	c_4 = \frac{s\, \mathcal{N}_4}{\mathcal{D}},
	\label{eq:correction-factors}
\end{equation}
where the numerators \(\mathcal{N}_i\) are \(3\times 3\) determinants built from the spatial components of the four-momenta, and the common denominator \(\mathcal{D}\) is a \(4\times 4\) determinant involving both the energies \(E_i\) and the spatial components (see Appendix~\ref{app:correction}). The factors \(c_i\) represent the relative energy-scale corrections needed to bring the measured \(b\)-jets and top quarks onto the mass shell of the heavy resonance. Events with any \(c_i \le 0\) are rejected.	
	\section{Top-Tagging Performance and Event Kinematics}
	\label{sec:top-tagging}
	
	Before presenting the final results, it is instructive to validate the top-tagging strategy and to quantify whether jet substructure provides any discrimination between the signal (\(H^\pm \to tb\)) and the dominant \(t\bar{t}\) background. All large-\(R\) jets are clustered with \(R=2.0\) and the \(N\)-subjettiness variables \(\tau_1,\dots,\tau_4\) are computed.
	
\begin{figure}[t]
		\centering
		\includegraphics[width=0.48\linewidth]{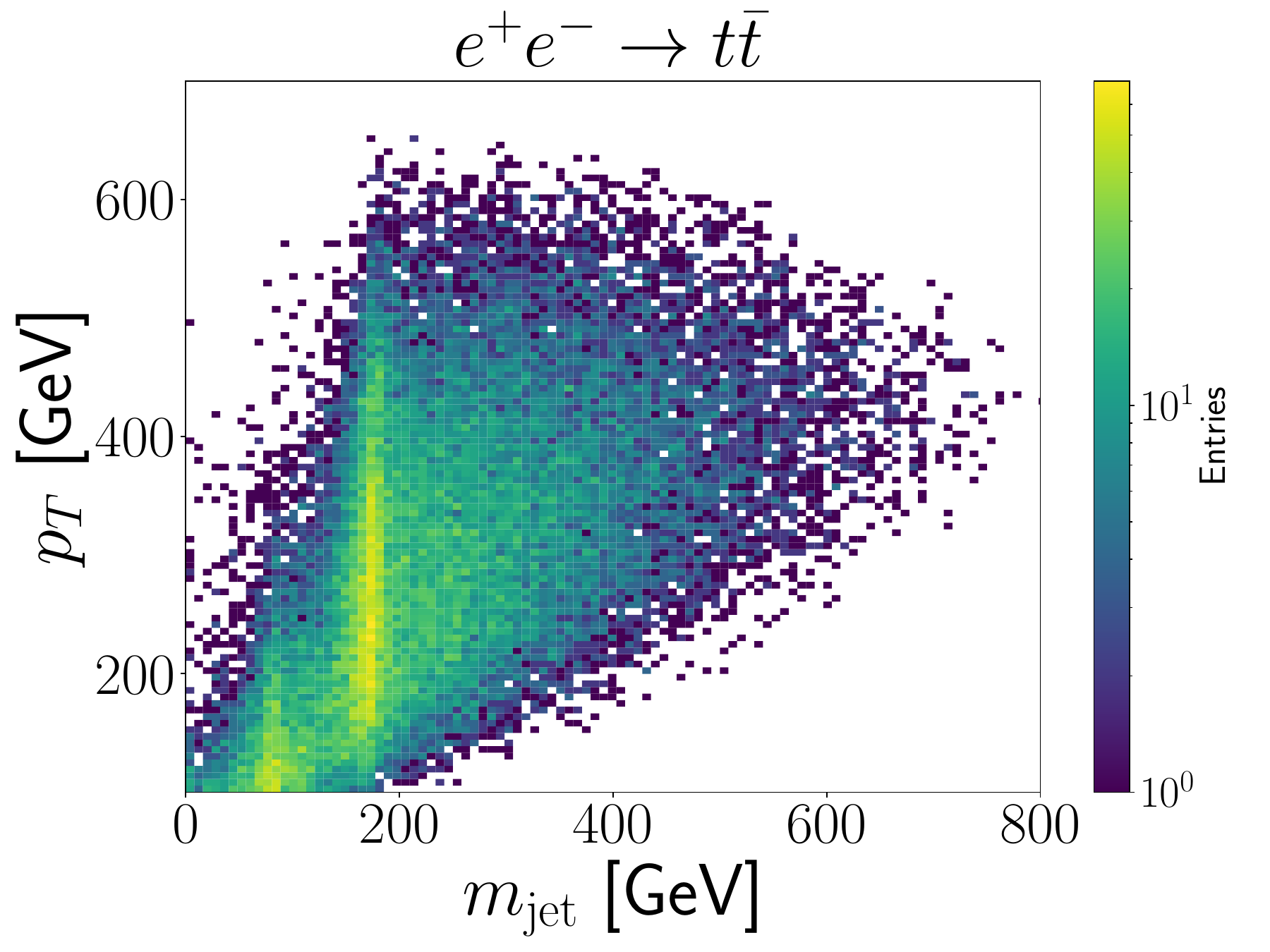}
		\includegraphics[width=0.48\linewidth]{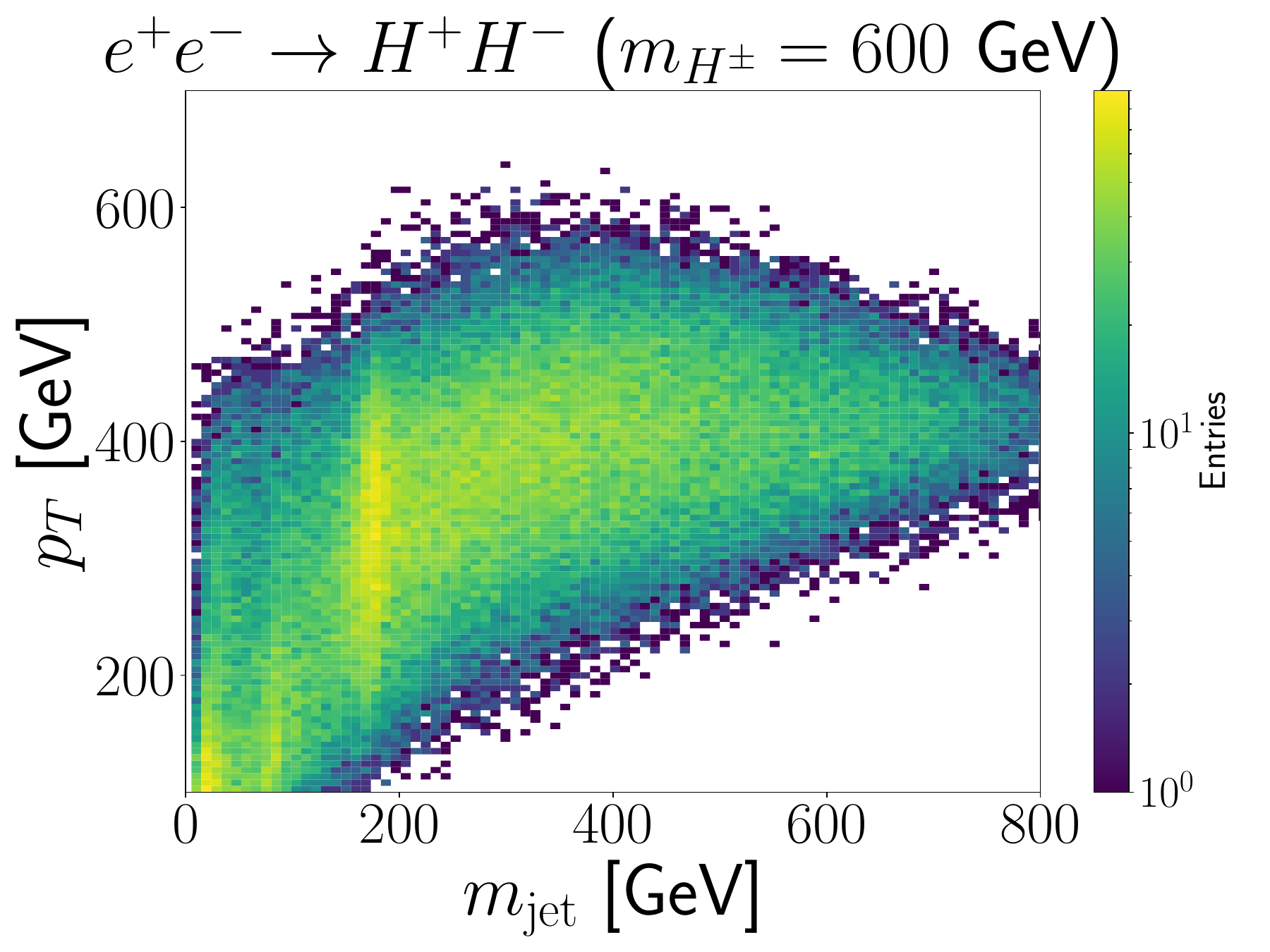}
		\caption{Large-\(R\) jet mass \(m_{\rm jet}\) versus jet \(\pt\) for \(e^+e^- \to t\bar{t}\) (left) and \(e^+e^- \to H^+H^-\) with \(m_{H^\pm}=600~\GeV\) (right). Both samples show a top-mass peak near \(m_t\approx 173~\GeV\); the signal tops are visibly harder.}
		\label{fig:mjet_pt}
	\end{figure}

	Figure~\ref{fig:mjet_pt} shows the two-dimensional distribution of the large-\(R\) jet mass \(m_{\rm jet}\) versus jet \(\pt\) for the \(t\bar{t}\) sample and for the \(H^\pm H^\mp\) sample with \(m_{H^\pm}=600~\GeV\). In both cases the jet mass peaks near \(m_t\approx 173~\GeV\), confirming that the tops are indeed captured in a single large-\(R\) jet. 
	
\begin{figure}[t]
		\centering
		\includegraphics[width=0.48\linewidth]{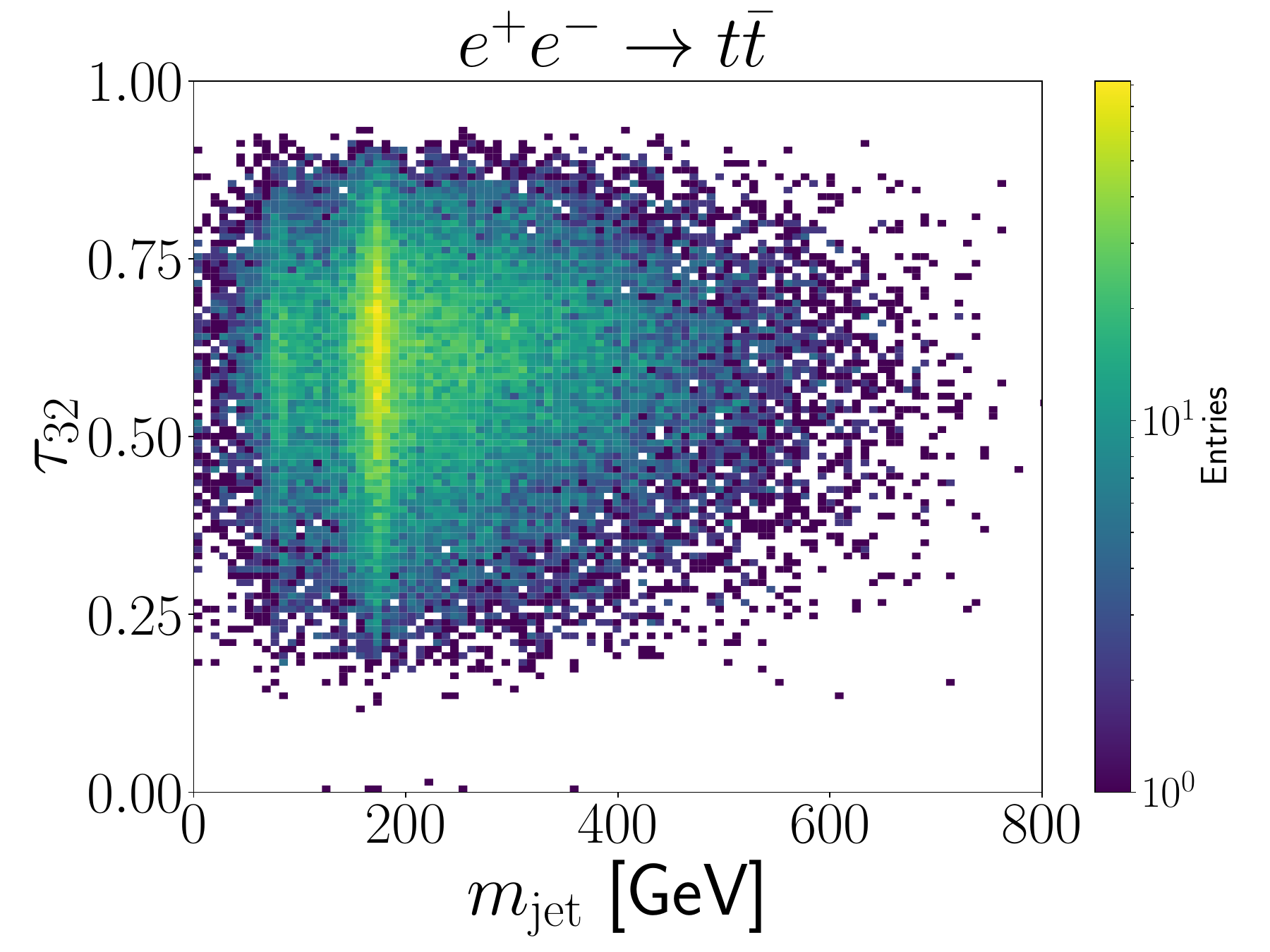}
		\includegraphics[width=0.48\linewidth]{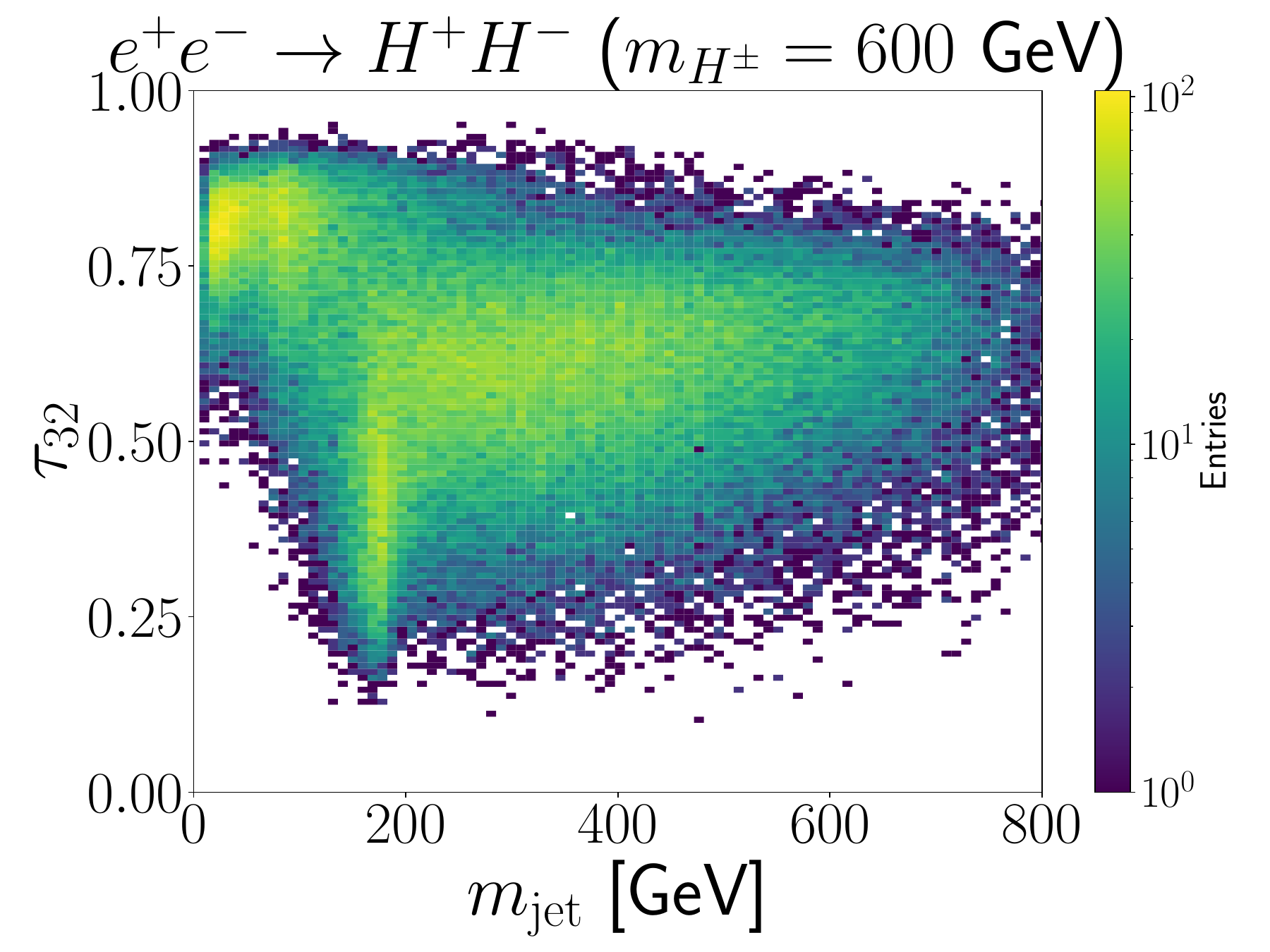}
		\caption{Large-\(R\) jet mass \(m_{\rm jet}\) versus \(\tau_{32}\) for \(t\bar{t}\) (left) and \(H^\pm H^\mp\) with \(m_{H^\pm}=600~\GeV\) (right). The top-mass band and its \(\tau_{32}\) spread overlap almost completely between the two samples.}
		\label{fig:mjet_t32}
	\end{figure}

	Figure~\ref{fig:mjet_t32} shows the corresponding \(m_{\rm jet}\) versus \(\tau_{32}=\tau_3/\tau_2\) plane, the classic top-tagging working-point plane. Genuine top jets populate a band around \(m_{\rm jet}\approx m_t\) with \(\tau_{32}\sim 0.25\)--\(0.75\). The two samples overlap almost completely in this plane: the top-mass band and its \(\tau_{32}\) spread are similar in both cases.
	
\begin{figure}[t]
		\centering
		\includegraphics[width=0.48\linewidth]{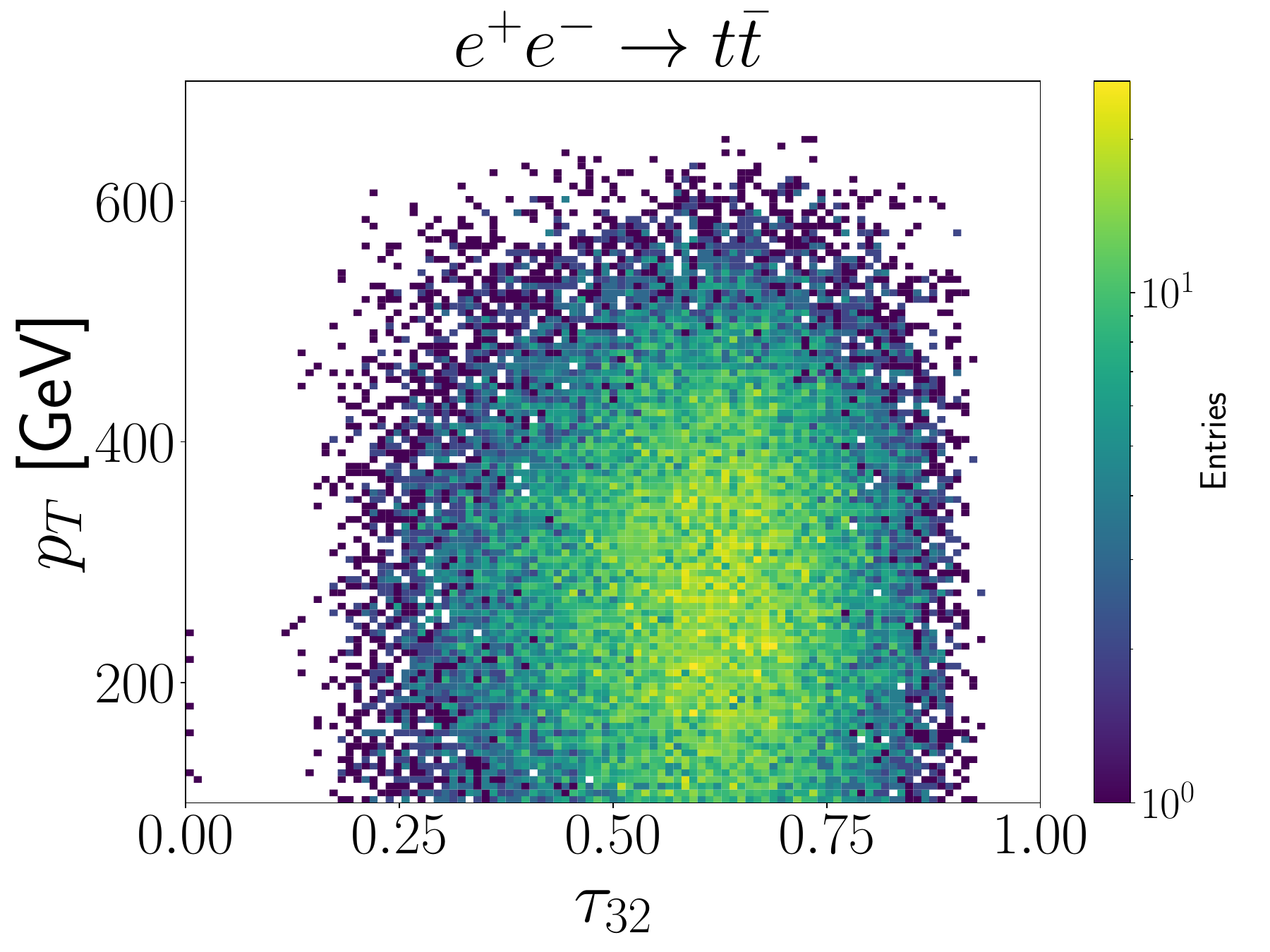}
		\includegraphics[width=0.48\linewidth]{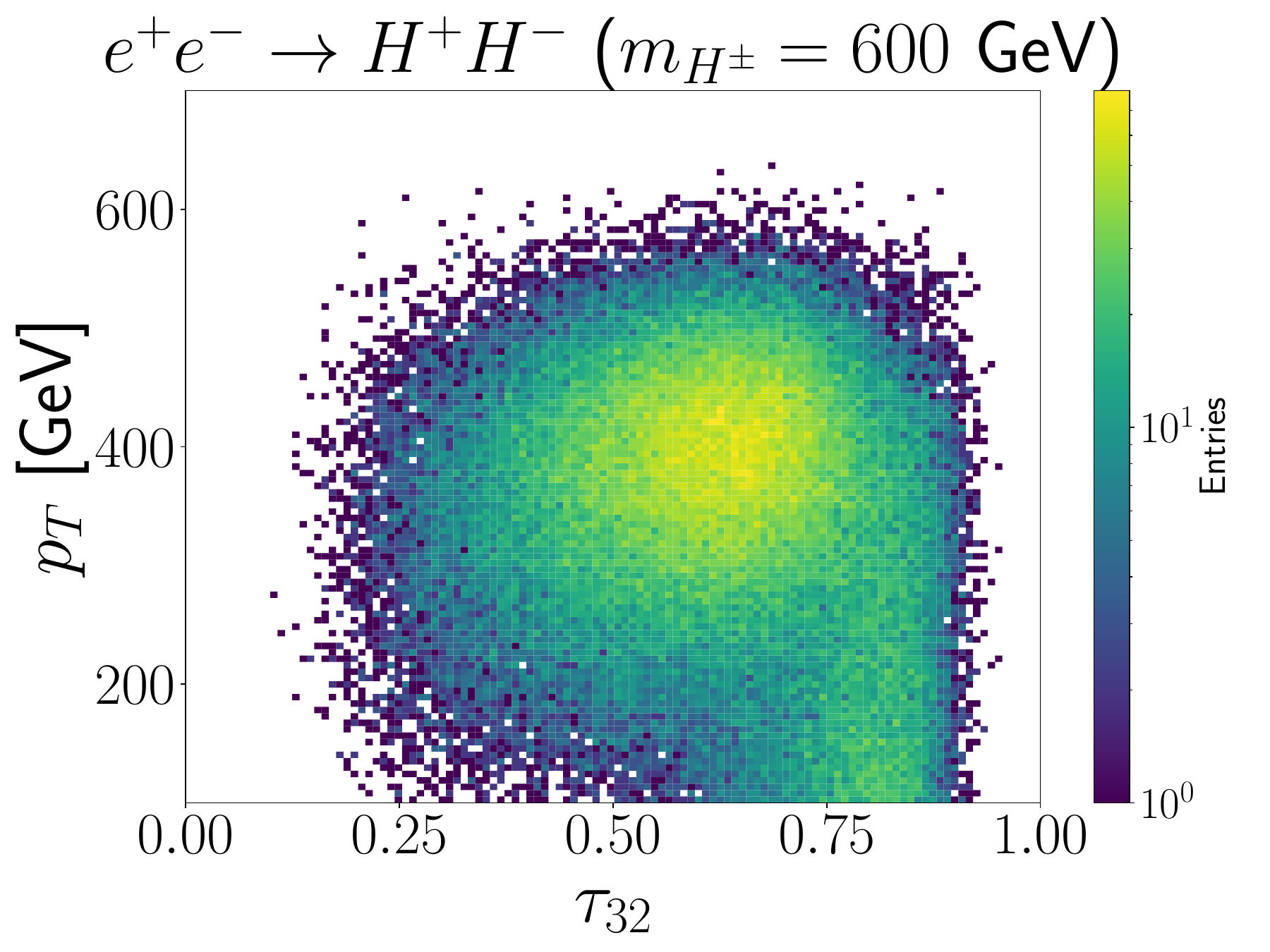}
		\caption{\(\tau_{32}\) versus jet \(\pt\) for \(t\bar{t}\) (left) and \(H^\pm H^\mp\) with \(m_{H^\pm}=600~\GeV\) (right). The correlation between \(\tau_{32}\) and \(\pt\) is the same in both samples.}
		\label{fig:t32_pt}
	\end{figure}

	To isolate the substructure from the kinematic difference, Fig.~\ref{fig:t32_pt} shows \(\tau_{32}\) versus \(\pt\). No sizable correlation is observed except for the signal which shows that as \(\pt\) increases, the prongs merge and \(\tau_{32}\) slightly decreases.
	
\begin{figure}[t]
		\centering
		\includegraphics[width=0.8\linewidth]{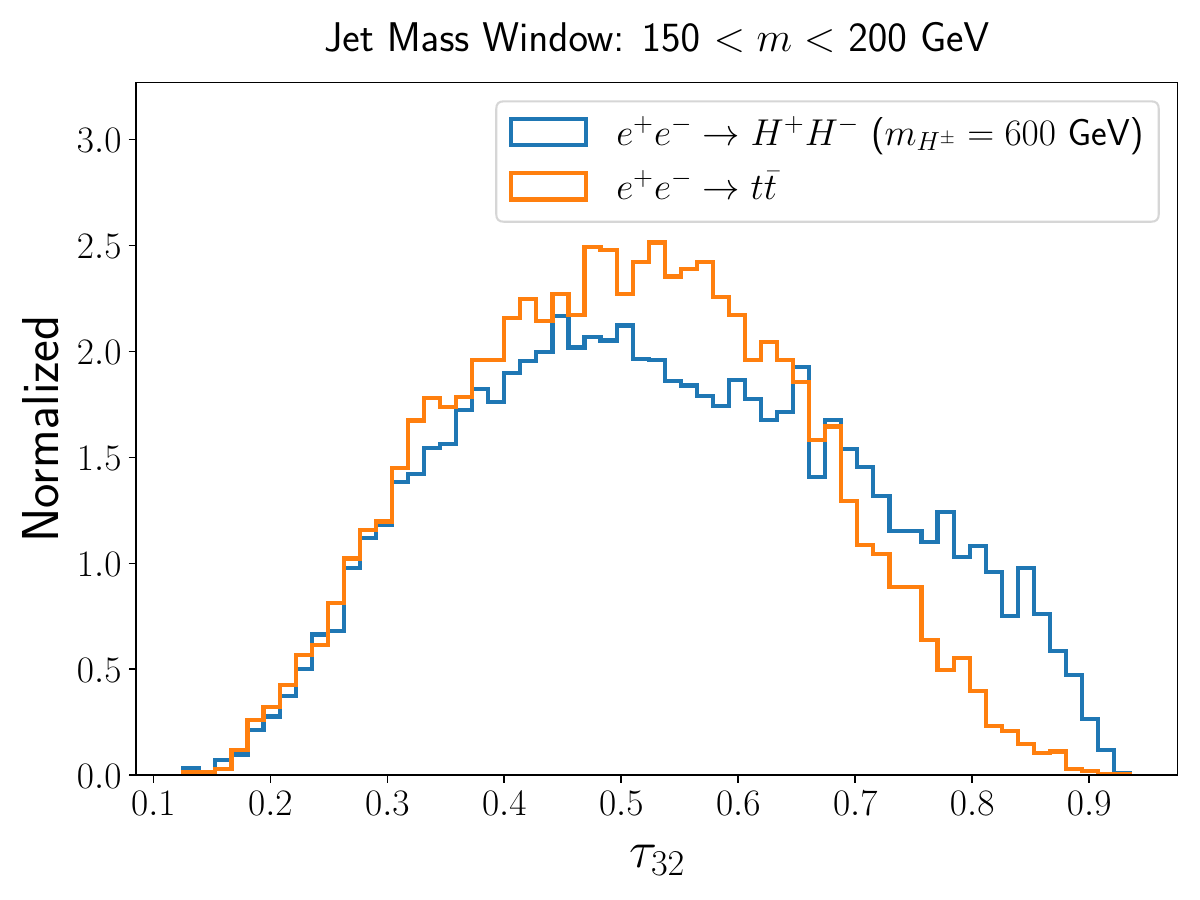}
		\caption{Normalized \(\tau_{32}\) distribution for \(H^\pm \to tb\) (\(m_{H^\pm}=600~\GeV\)) and prompt \(t\bar{t}\), restricted to the top-jet mass window \(150 < m_{\rm jet} < 200~\GeV\). The two distributions overlap, confirming that top-tagging substructure does not discriminate the two production mechanisms.}
		\label{fig:t32_overlay}
	\end{figure}

	Figure~\ref{fig:t32_overlay} makes the point quantitative. It shows the normalized \(\tau_{32}\) distribution for both samples in a common jet-mass window, \(150 < m_{\rm jet} < 200~\GeV\), which selects genuine top jets. The two distributions lie on top of each other: any apparent difference in the inclusive \(\tau_{32}\) spectrum is a \(\pt\) effect, not a difference in top substructure. It is therefore concluded that jet substructure observables do \emph{not} provide a handle for separating \(H^\pm \to tb\) from prompt \(t\bar{t}\) at CLIC energies. This is expected: the top decay \(t\to Wb\) is the same regardless of the production mechanism, the initial state is electroweak and colorless (so there is no color-flow or underlying-event imprint), and the jet substructure is boost-invariant up to the kinematic difference already discussed.
	
\begin{figure}[t]
		\centering
		\includegraphics[width=0.48\linewidth]{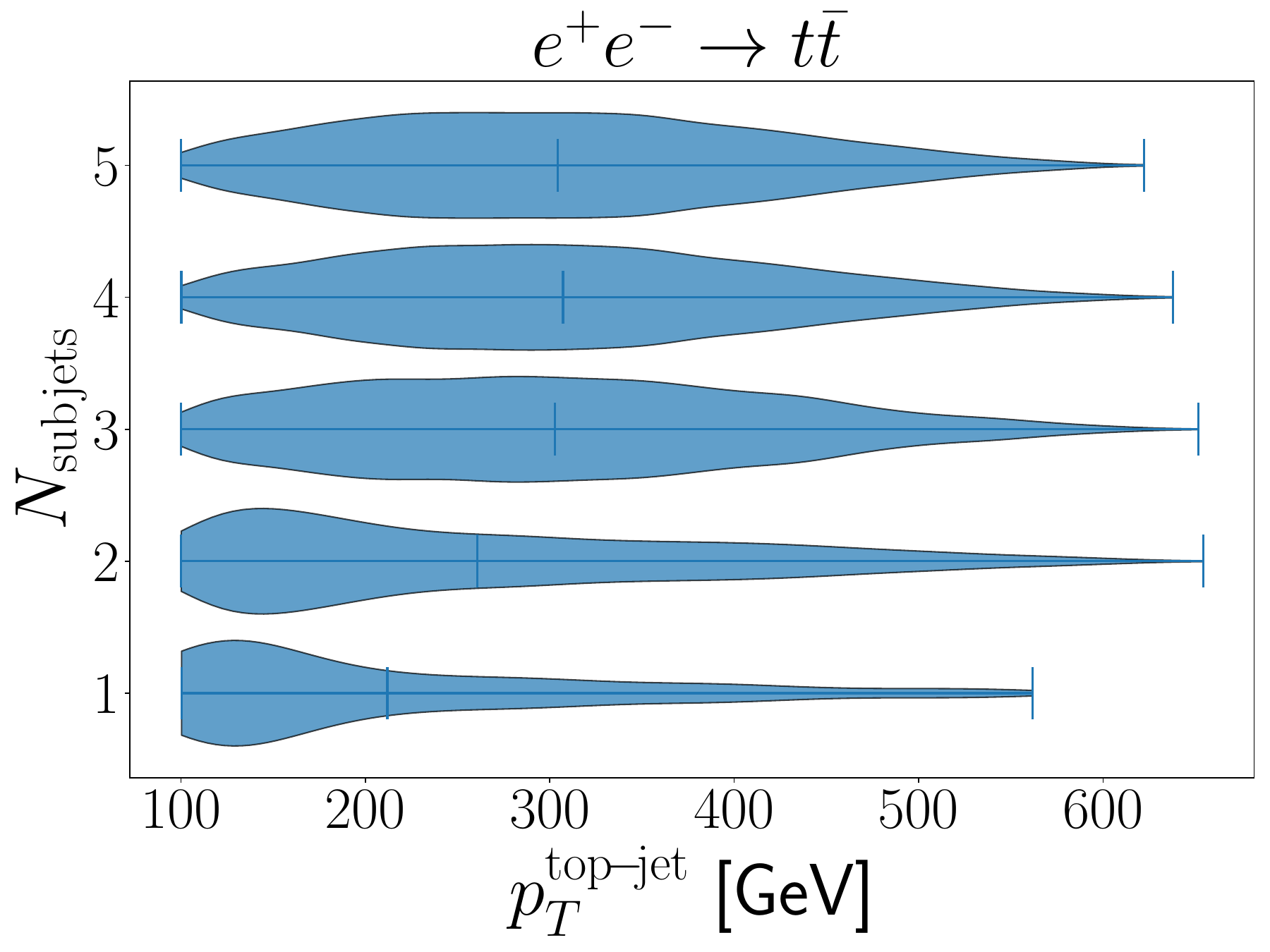}
		\includegraphics[width=0.48\linewidth]{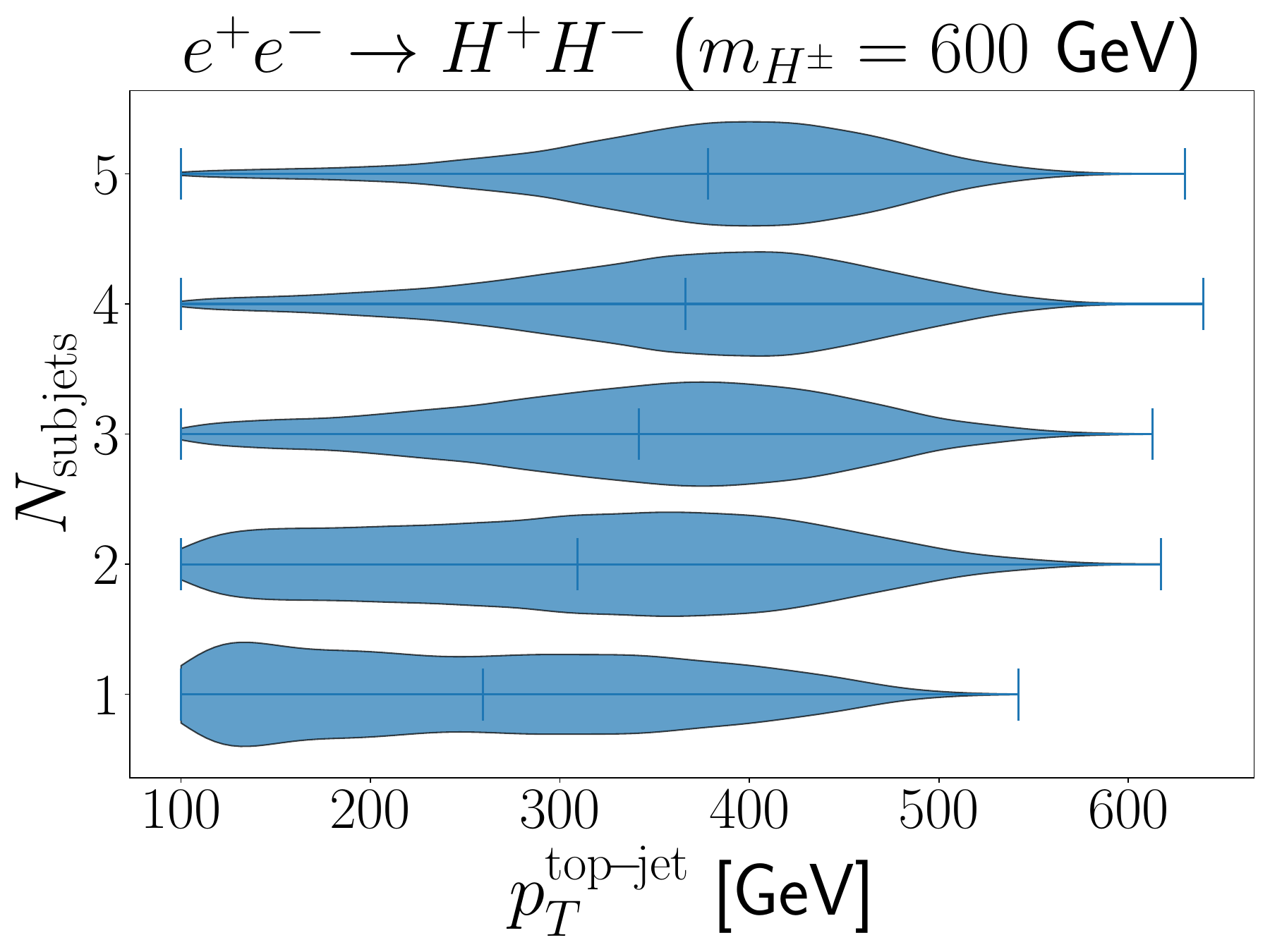}
		\caption{Violin plots of the number of subjets \(N_{\text{subjets}}\) versus the jet transverse momentum \(\pt^{\text{top-jet}}\), for the \(t\bar{t}\) background (left) and the \(H^\pm H^\mp\) signal with \(m_{H^\pm}=600~\GeV\) (right). The distributions are similar in shape, confirming that the top-tagging variables do not separate the two samples.}
		\label{fig:violin}
	\end{figure}

	To further validate the top-tagging procedure and the behaviour of the large-\(R\) jets, the distribution of the number of subjets \(N_{\text{subjets}}\) versus the jet transverse momentum \(\pt^{\text{top-jet}}\) is shown in Fig.~\ref{fig:violin}, for the \(t\bar{t}\) background (top) and the \(H^\pm H^\mp\) signal with \(m_{H^\pm}=600~\GeV\) (bottom). The distributions are presented as violin plots, showing the probability density of the subjet multiplicity in bins of \(\pt\). The differences between the signal and background distributions are largely associated with the different \(\pt\) spectra of the top quarks. In both samples, the subjet multiplicity tends to increase with increasing top-quark \(\pt\), and the two distributions become increasingly similar in the high-\(\pt\) region, where both exhibit sufficient subjet multiplicity for top identification. Consequently, the subjet multiplicity alone provides limited discrimination between signal and background, and the separation must rely primarily on event-level kinematic observables and the \(b\)-jet multiplicity, as discussed below.
	
	\subsection{Object Multiplicities}
	
	The discrimination between the signal and the dominant backgrounds relies heavily on event-level object multiplicities. The signal process \(H^+H^- \to (tb)(tb)\) yields a \(t\bar{t}b\bar{b}\) final state, which naturally contains four \(b\)-quarks. In contrast, the dominant \(t\bar{t}\) background contains only two \(b\)-quarks; any additional \(b\)-jets arise from hard gluon splitting or mis-tagging.
	
\begin{figure}[t]
		\centering
		\includegraphics[width=0.9\linewidth]{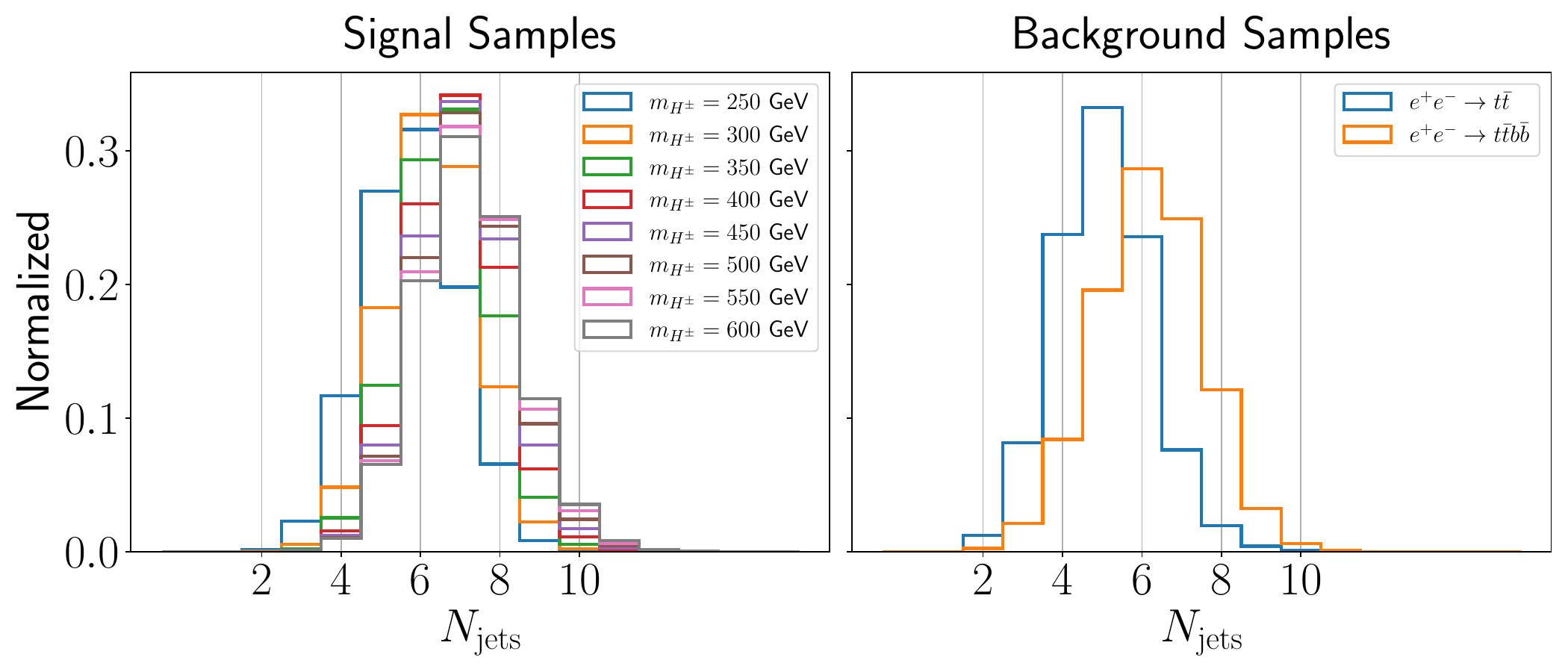}\\[2mm]
		\includegraphics[width=0.9\linewidth]{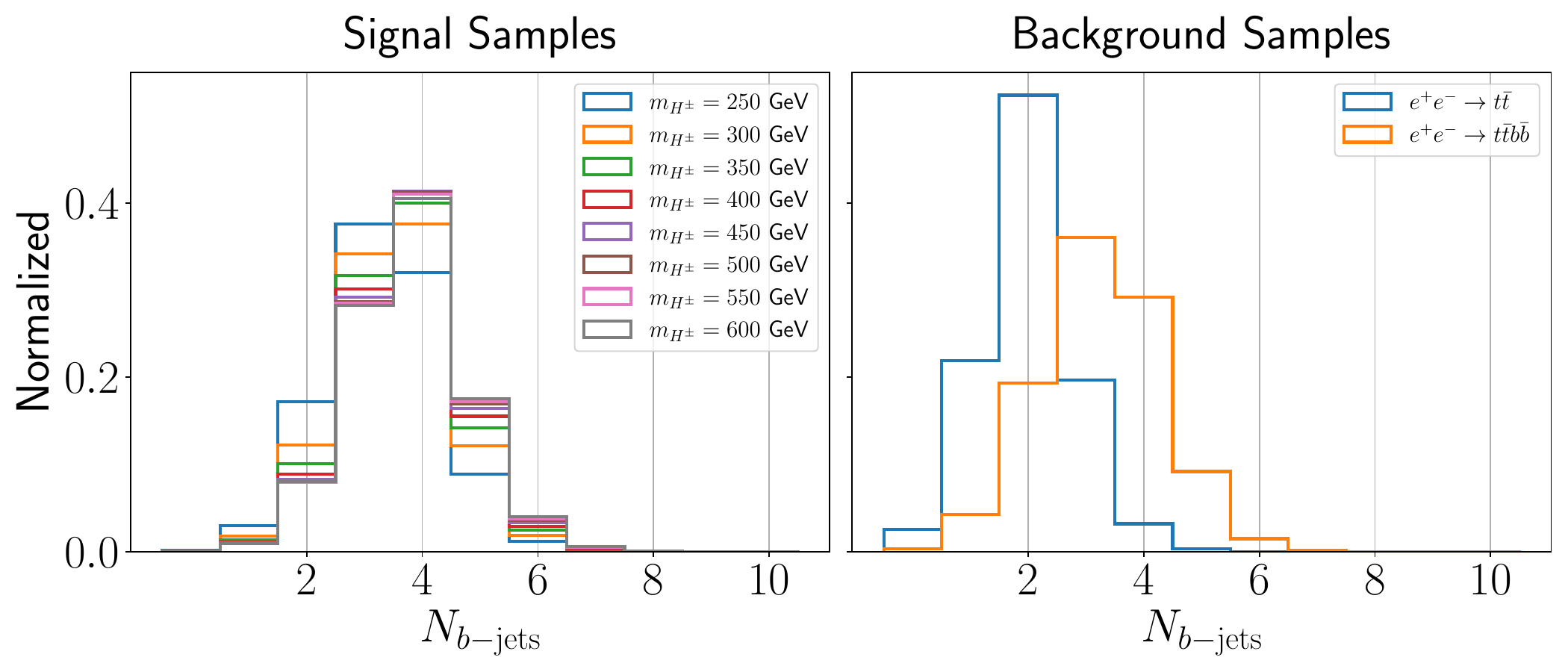}\\[2mm]
		\includegraphics[width=0.9\linewidth]{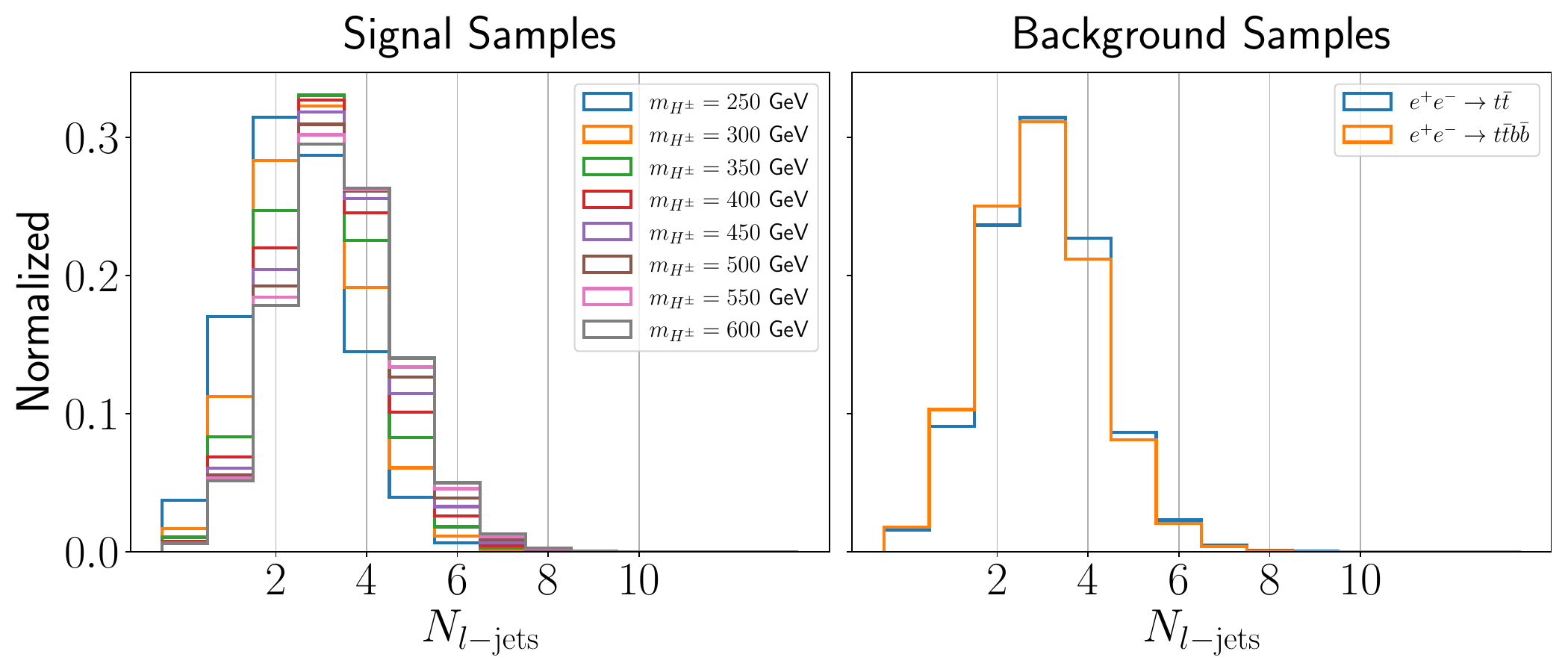}\\[2mm]
		\includegraphics[width=0.9\linewidth]{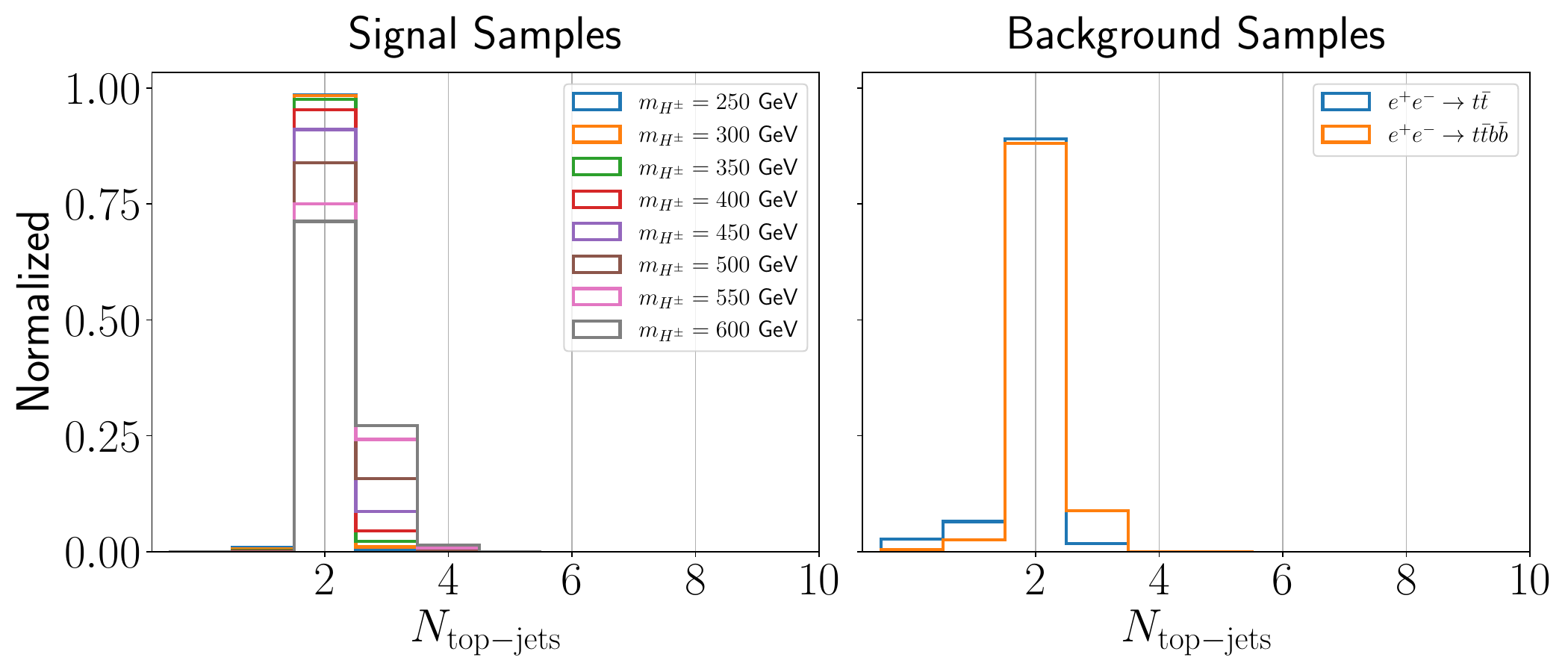}
		\caption{Normalized distributions of the number of jets (row 1), \(b\)-jets (row 2), light jets (row 3), and top-tagged jets (row 4). In each row, the signal (multiple masses, left) is compared with the \(t\bar{t}\) background (right). The signal shows a clear peak at four \(b\)-jets, whereas the \(t\bar{t}\) background is heavily suppressed in that region.}
		\label{fig:multiplicities}
	\end{figure}

	Figure~\ref{fig:multiplicities} shows the normalized distributions of the number of jets (\(N_{\text{jets}}\)), \(b\)-jets (\(N_{b\text{-jets}}\)), light jets (\(N_{\ell\text{-jets}}\)), and top-tagged jets (\(N_{\text{top-jets}}\)) for the signal (left panel) and background (right panel). While the signal and background have similar distributions for \(N_{\text{jets}}\) and \(N_{\text{top-jets}}\), they are strikingly different for \(N_{b\text{-jets}}\). The signal peaks at four \(b\)-jets, while the background drops sharply. This justifies the \(\ge 4\) \(b\)-jet requirement, which effectively eliminates the \(t\bar{t}\) background (reducing its efficiency to \(0.036\%\), as shown in Table~\ref{tab:cutflow}), while retaining a substantial fraction of the signal.
	
\begin{figure}[t]
		\centering
		\includegraphics[width=0.9\linewidth]{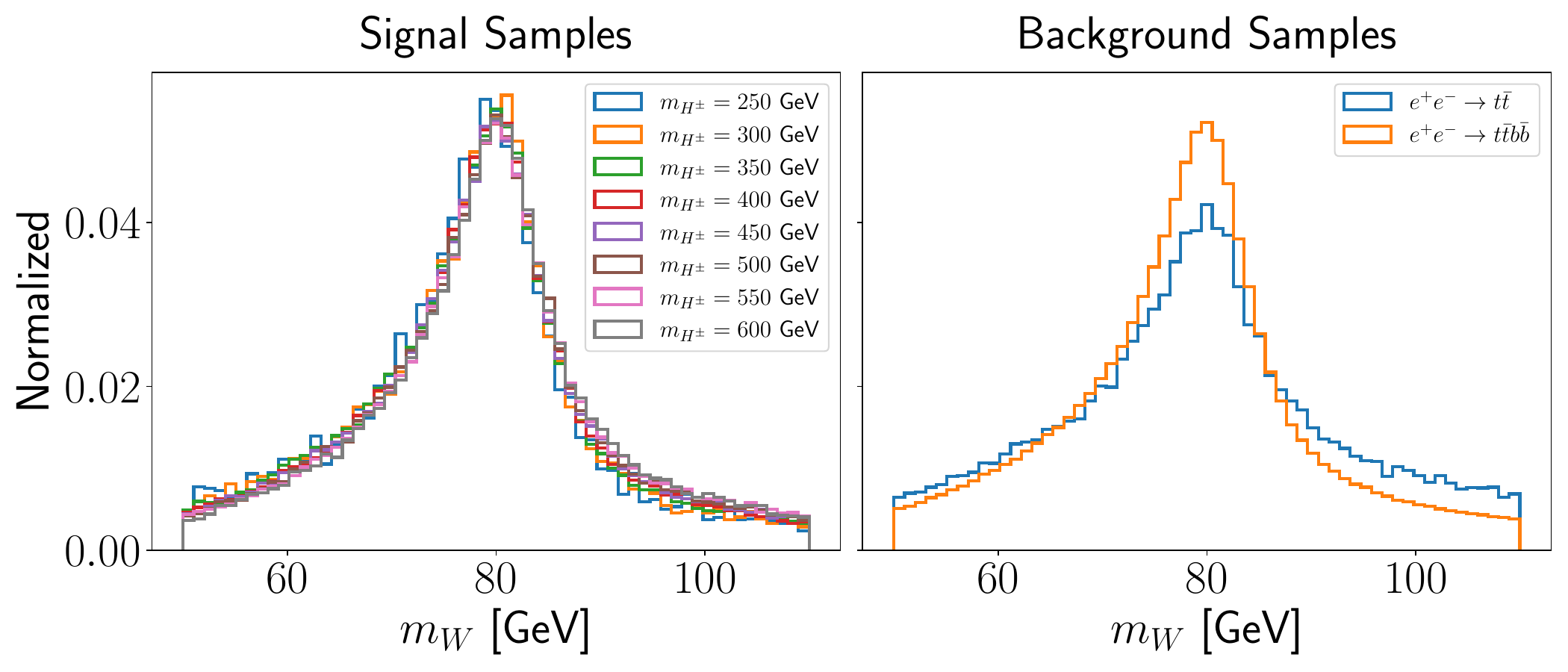}\\[2mm]
		\includegraphics[width=0.9\linewidth]{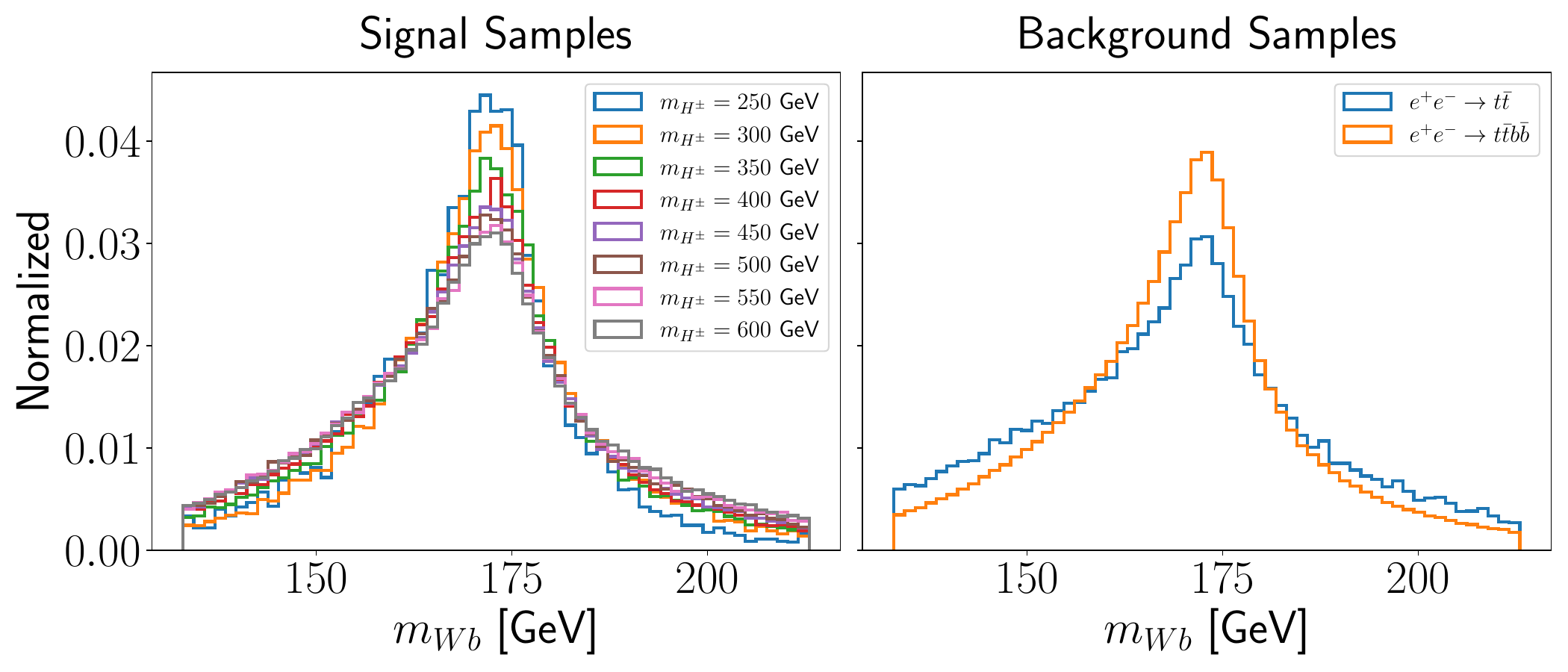}
		\caption{Reconstructed \(W\)-boson mass \(m_W\) (top) and \(Wb\) mass \(m_{Wb}\) (bottom), with the two candidates per event concatenated. The signal (left) and background (right) both peak at the expected values; these windows are used for jet assignment, not for signal discrimination.}
		\label{fig:mW_mWb}
	\end{figure}

	Figure~\ref{fig:mW_mWb} shows the reconstructed \(W\)-boson and \(Wb\) (top-candidate) masses, where the two \(W\)'s and two \(Wb\)'s per event are concatenated. The signal and background distributions peak at \(m_W\) and \(m_t\) respectively, as expected from the reconstruction procedure; these windows are used to select the correct jet assignments, not to discriminate signal from background.
	
	\section{Results}
	\label{sec:results}
	
	The expected selection efficiencies at each stage of the analysis for the signal \(m_{H^\pm} = 600~\GeV\) and the dominant \(t\bar{t}\) background are summarized in Table~\ref{tab:cutflow}. The selection requirements successfully reduce the background while retaining a substantial fraction of the signal.
	
	\begin{table}[b]
		\caption{Cut-flow efficiencies for the signal (\(m_{H^\pm}=600~\GeV\)) and the \(t\bar{t}\) background.}
		\label{tab:cutflow}
		\begin{ruledtabular}
			\begin{tabular}{lcc}
				Selection stage & Signal Efficiency & \(t\bar{t}\) Efficiency \\
				\hline
				\(\geq 2\) top jets        & 99.98\% & 90.83\% \\
				\(\geq 4\) \(b\)-jets        & 62.64\% & 3.57\% \\
				\(\geq 4\) light jets      & 35.48\% & 8.63\% \\
				\(W\) mass window          & 65.01\% & 50.29\% \\
				Top mass window          & 78.66\% & 79.83\% \\
				Kinematic correction     & 72.91\% & 31.81\% \\
				\hline
				Total                    & 8.28\% & 0.036\% \\
			\end{tabular}
		\end{ruledtabular}
	\end{table}
	
\begin{figure}[t]
		\centering
		\includegraphics[width=0.9\linewidth]{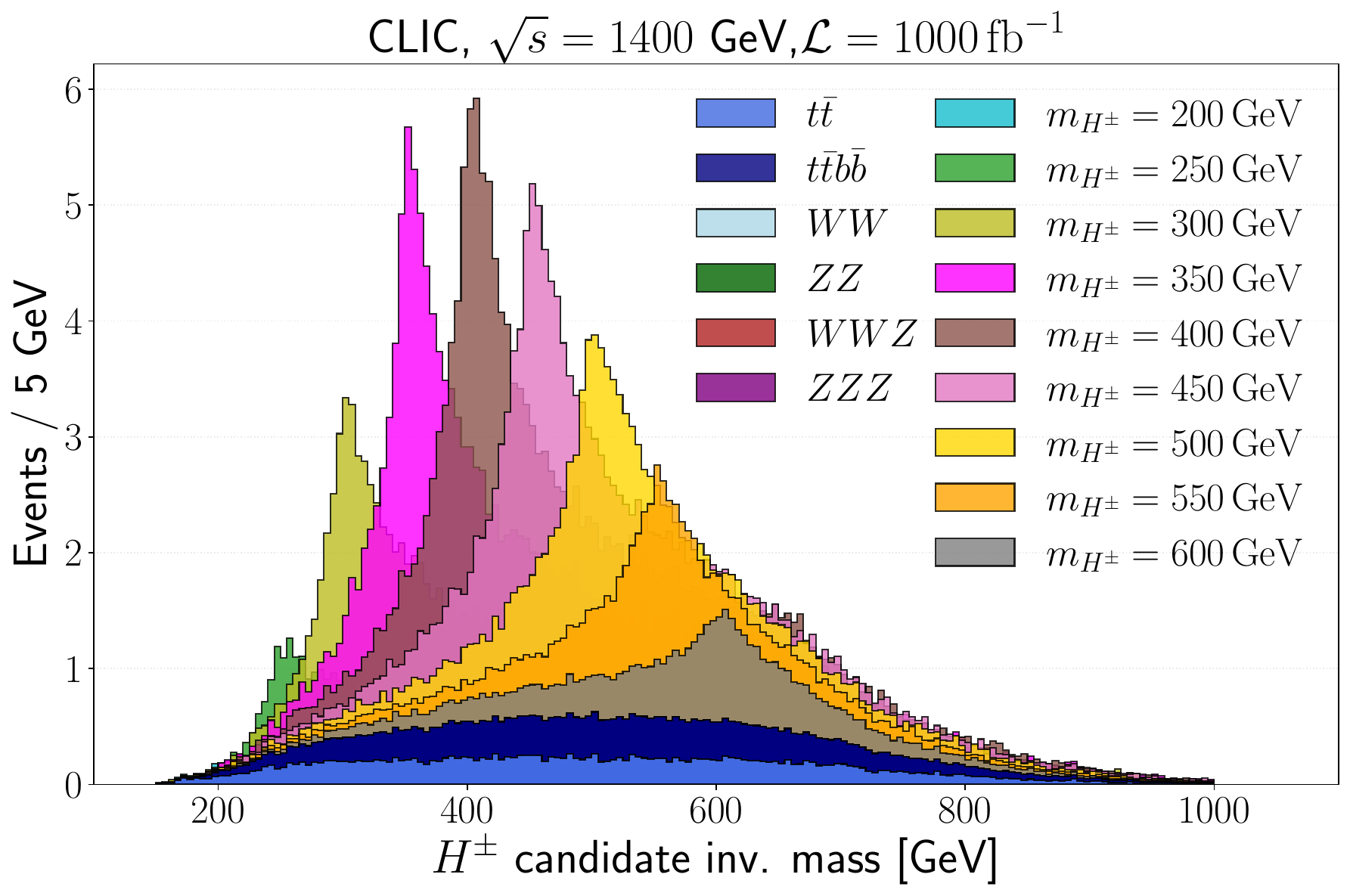}
		\caption{Reconstructed charged-Higgs mass \(m_{CH}\) for the signal mass points (colored) and the sum of backgrounds (filled). The signal forms a peak at the input \(m_{H^\pm}\), while the background is a broad continuum.}
		\label{fig:mCH}
	\end{figure}

	The reconstructed charged Higgs mass distribution (Fig.~\ref{fig:mCH}) exhibits a clear peak at the true charged Higgs mass for the signal, while the background remains a broad continuum.
	
The expected signal (\(N_S\)) and background (\(N_B\)) yields are evaluated in the optimal mass window for each charged Higgs mass hypothesis. A 10\% background systematic uncertainty is assumed through the statistical significance computed using the profiled Asimov formula:
\begin{equation}
	\begin{aligned}
		Z = \Bigg\{2\Bigg[ &(N_S+N_B)
		\ln\left(
		\frac{(N_S+N_B)(N_B+\sigma_B^2)}
		{N_B^2+(N_S+N_B)\sigma_B^2}
		\right)\\
		&-\frac{N_B^2}{\sigma_B^2}
		\ln\left(
		1+\frac{\sigma_B^2N_S}
		{N_B(N_B+\sigma_B^2)}
		\right)
		\Bigg]\Bigg\}^{1/2}.
	\end{aligned}
	\label{eq:asimov}
\end{equation}
where \(\sigma_B = 0.10 \times N_B\) is the absolute background uncertainty. This profile-likelihood approach provides a more realistic estimate of the discovery potential than the simple \(S/\sqrt{B}\) approximation, particularly in regions with low background yields.

The results of the cut-based analysis for the Type-I scenario at \(\tan\beta = 5\) are summarized in Table~\ref{tab:significance}. The significance peaks at \(m_{H^\pm} \approx 400\)~GeV, reaching \(Z \approx 12.2\), and remains above \(5\sigma\) for masses between roughly \(250\) and \(600\)~GeV.

\begin{table}[b]
	\caption{Expected signal (\(N_S\)) and background (\(N_B\)) yields, and the resulting profiled Asimov significance for the cut-based analysis with \(\mathcal{L} = 1000~\text{fb}^{-1}\). The background uncertainty is taken as 10\%.}
	\label{tab:significance}
	\begin{ruledtabular}
		\begin{tabular}{cccc}
			\(m_{H^\pm}\) (GeV) & \(N_S\) & \(N_B\) & \(Z_{\text{cut}}\) \\
			\hline
			200 & 0.50 & 1.64 & 0.37 \\
			250 & 13.05 & 8.64 & 3.51 \\
			300 & 47.49 & 14.29 & 8.27 \\
			350 & 80.00 & 16.81 & 11.56 \\
			400 & 96.49 & 21.04 & 12.22 \\
			450 & 107.81 & 30.17 & 11.51 \\
			500 & 83.21 & 27.88 & 9.78 \\
			550 & 48.68 & 23.42 & 6.91 \\
			600 & 24.53 & 25.74 & 3.71 \\
			\hline
		\end{tabular}
	\end{ruledtabular}
\end{table}
	
	\subsection{Branching ratios and the CLIC exclusion reach}
	\label{sec:br_excl}
	
\begin{figure}[t]
		\centering
		\includegraphics[width=0.95\linewidth]{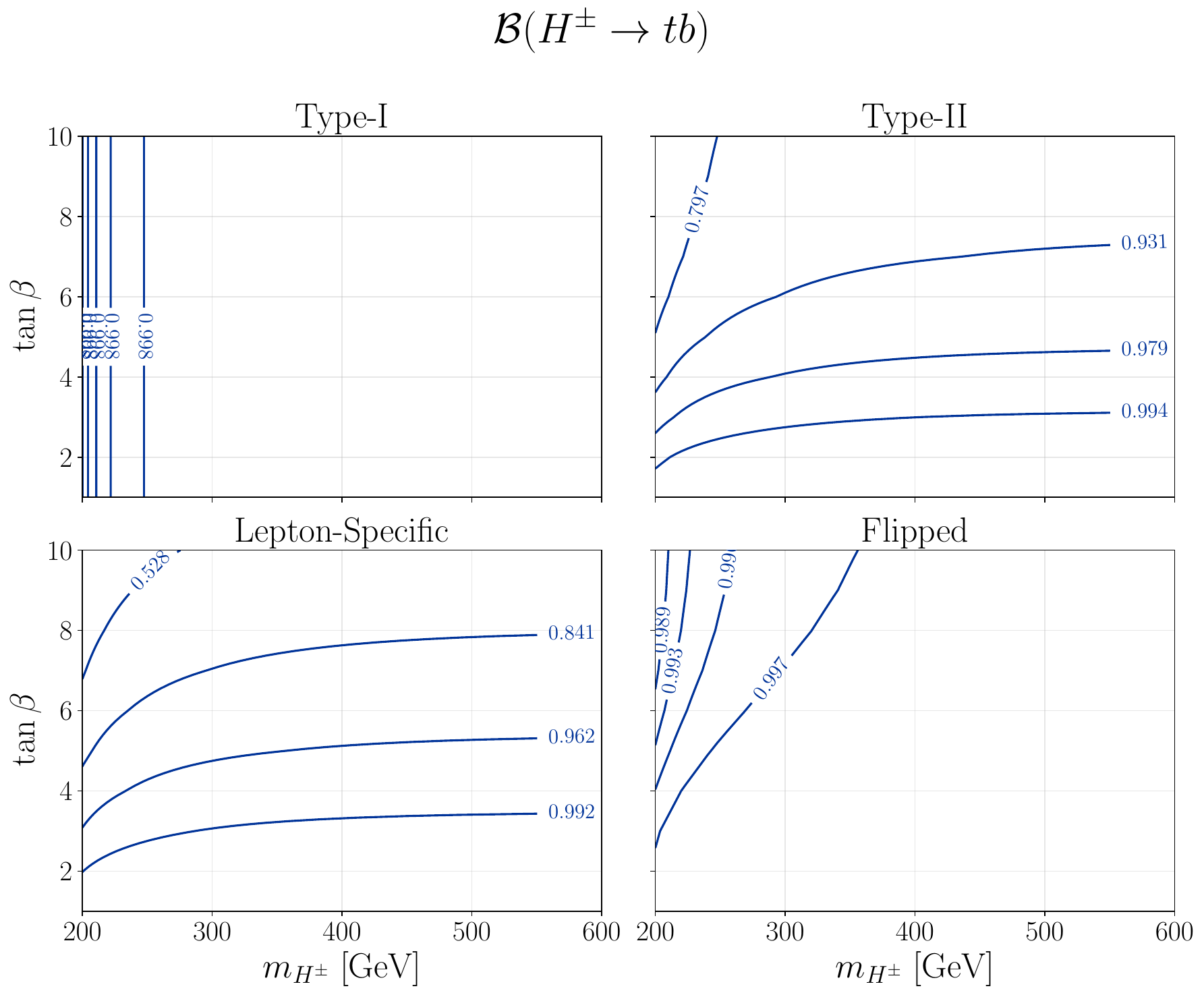}
		\caption{Branching ratio \(\mathcal{B}(H^\pm\to tb)\) in the \((m_{H^\pm},\tan\beta)\) plane for the four 2HDM types.}
		\label{fig:br_tb}
	\end{figure}

\begin{figure}[t]
		\centering
		\includegraphics[width=0.95\linewidth]{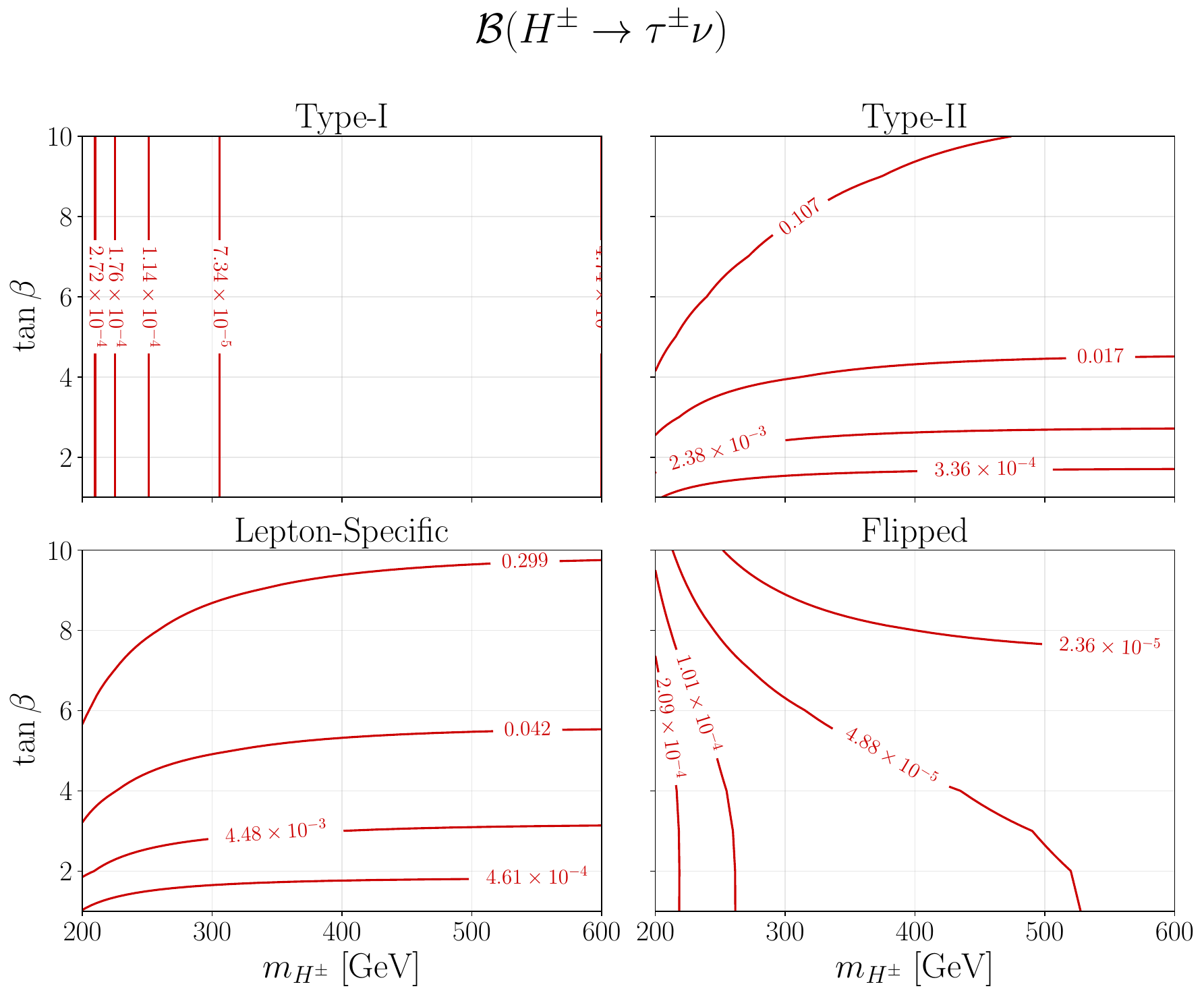}
		\caption{Branching ratio \(\mathcal{B}(H^\pm\to\tau^\pm\nu)\) in the same plane and for the same four types.}
		\label{fig:br_taunu}
	\end{figure}

	The CLIC reach in the \(H^\pm \to tb\) channel depends not only on the production cross section, which is model-independent, but also on the branching ratio \(\mathcal{B}(H^\pm \to tb)\), which is type-dependent. Figures~\ref{fig:br_tb} and~\ref{fig:br_taunu} show the two dominant decay branching ratios in the \((m_{H^\pm}, \tan\beta)\) plane for the four Yukawa types. The contour levels are drawn in logit space, \(\log[\mathcal{B}/(1-\mathcal{B})]\), so that values near 0 and near 1 are both visible.
	
Figure~\ref{fig:br_tb} shows the branching ratio of the charged Higgs boson into \(tb\) for the four Yukawa types. An important observation is that this decay remains dominant over almost the entire parameter space considered, with \(\mathcal{B}(H^\pm\to tb)\) close to unity. This is particularly evident in Type-I, where the branching ratio is nearly constant at \(\mathcal{B}\approx 0.997\)--\(0.998\). A similarly strong dominance is observed in the Lepton-Specific model, despite the \(\tan\beta\)-enhancement of the \(\tau\nu\) coupling. This behaviour can be understood from the large top-quark mass, together with the colour factor associated with the quark decay, which substantially enhances the \(tb\) partial width relative to the leptonic mode over the parameter range considered. The Type-II and Flipped models exhibit more pronounced variations, reflecting the different \(\tan\beta\) dependence of their fermionic couplings and the competition among the available decay channels.

The complementary branching ratios for \(H^\pm\to\tau\nu\) are shown in Fig.~\ref{fig:br_taunu}. They remain strongly suppressed over most of the scanned parameter space, consistent with the dominance of the \(tb\) decay mode.
	
Combining these branching ratios with the model-independent production cross section, the expected signal yield for the \(tb\) channel is
\[
\begin{split}
	N_S(m_{H^\pm}, \tan\beta) &= \sigma(e^+e^-\to H^+H^-) \\
	&\quad \times \mathcal{B}^2(H^\pm\to tb)\,
	\mathcal{B}^2(W \to jj)\,
	\epsilon\,
	\mathcal{L}_{\rm int}\,,
\end{split}
\]
where the square comes from the two charged Higgs bosons per event. The overall reconstruction and selection efficiency, \(\epsilon\), accounts for the detector acceptance, object reconstruction, and the specific kinematic and \(b\)-tagging requirements of the analysis. It is obtained from the Monte Carlo simulation as the fraction of generated signal events that pass the full event selection.

\begin{figure}[t]
		\centering
		\includegraphics[width=0.95\linewidth]{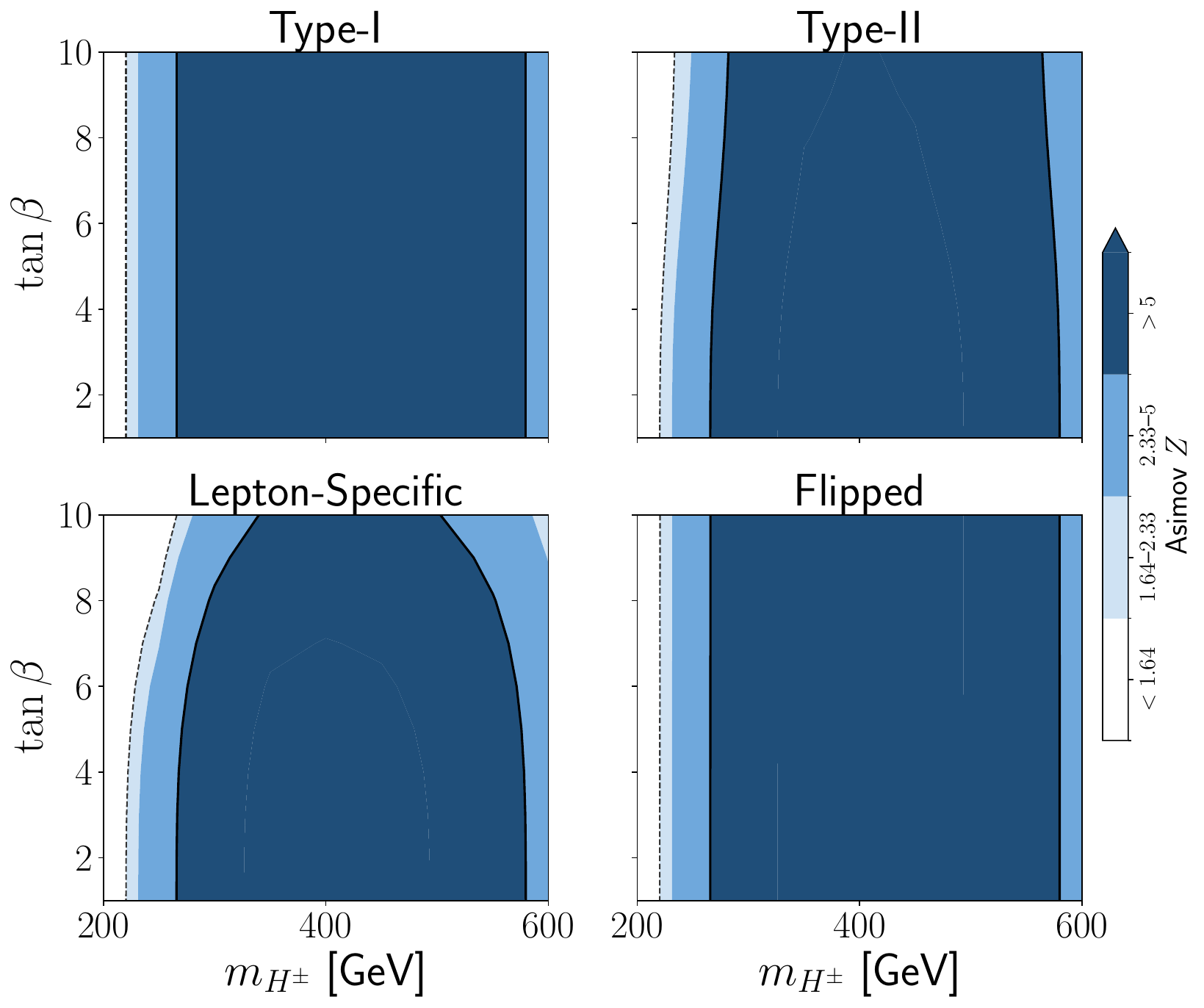}
		\caption{CLIC exclusion reach in the \(H^\pm \to tb\) channel for \(\mathcal{L}_{\rm int} = 1000~\text{fb}^{-1}\), in the \((m_{H^\pm}, \tan\beta)\) plane for the four 2HDM types. Shading: Asimov \(Z > 1.64\) (\(95\%\) CL, light blue), \(Z > 2.33\) (\(99\%\) CL, medium blue), \(Z > 5\) (dark blue). The dashed line marks the \(Z = 1.64\) boundary.}
		\label{fig:clic_excl}
	\end{figure}

The CLIC exclusion reach in the \(H^\pm \to tb\) channel is shown in Fig.~\ref{fig:clic_excl}. It is computed using the Asimov \(Z\) in the optimal charged Higgs mass window, with a \(10\%\) background normalization uncertainty.
	
Since the pair-production cross section \(e^+e^-\to H^+H^-\) is independent of the Yukawa type and \(\tan\beta\), the differences in the CLIC exclusion contours in Fig.~\ref{fig:clic_excl} arise primarily from the model dependence of the charged Higgs branching ratios. For the \(tb\)-based analysis, the Type-I and Flipped models exhibit very similar behaviour, with a nearly vertical exclusion region extending over almost the entire \(\tan\beta\) range considered. A similarly broad reach is obtained in Type-II, although with some dependence on \(\tan\beta\). In contrast, the Lepton-Specific model shows a pronounced reduction in sensitivity at large \(\tan\beta\), where the increasing \(\tau\nu\) branching ratio reduces the fraction of signal events entering the \(tb\)-based analysis. Consequently, the CLIC reach becomes narrower in this region. This illustrates the complementarity between the \(tb\) and \(\tau\nu\) final states: while the latter can provide additional sensitivity at large \(\tan\beta\), particularly in the Lepton-Specific model, it simultaneously reduces the signal yield available to the \(tb\) channel. A combined analysis of both final states at CLIC could therefore improve the overall coverage of the parameter space.

	\section{Machine Learning Analysis}
	\label{sec:ml_analysis}
	
	To complement the cut-based search, a multivariate analysis using Boosted Decision Trees (BDTs) was performed. The BDT is trained to separate the charged Higgs signal from the dominant \ttbar, \ttbb, \ww, \zz, \wwz, and \zzz ~ backgrounds. The training uses a set of kinematic and topological features, including the reconstructed masses of the $W$ bosons and $b$-jets, jet multiplicities, and the kinematics of the top quarks. The most discriminating feature for the signal extraction is the reconstructed charged Higgs mass, $m_{H^\pm}$.
	
\begin{figure}[htbp]
		\centering
		\includegraphics[width=0.48\textwidth]{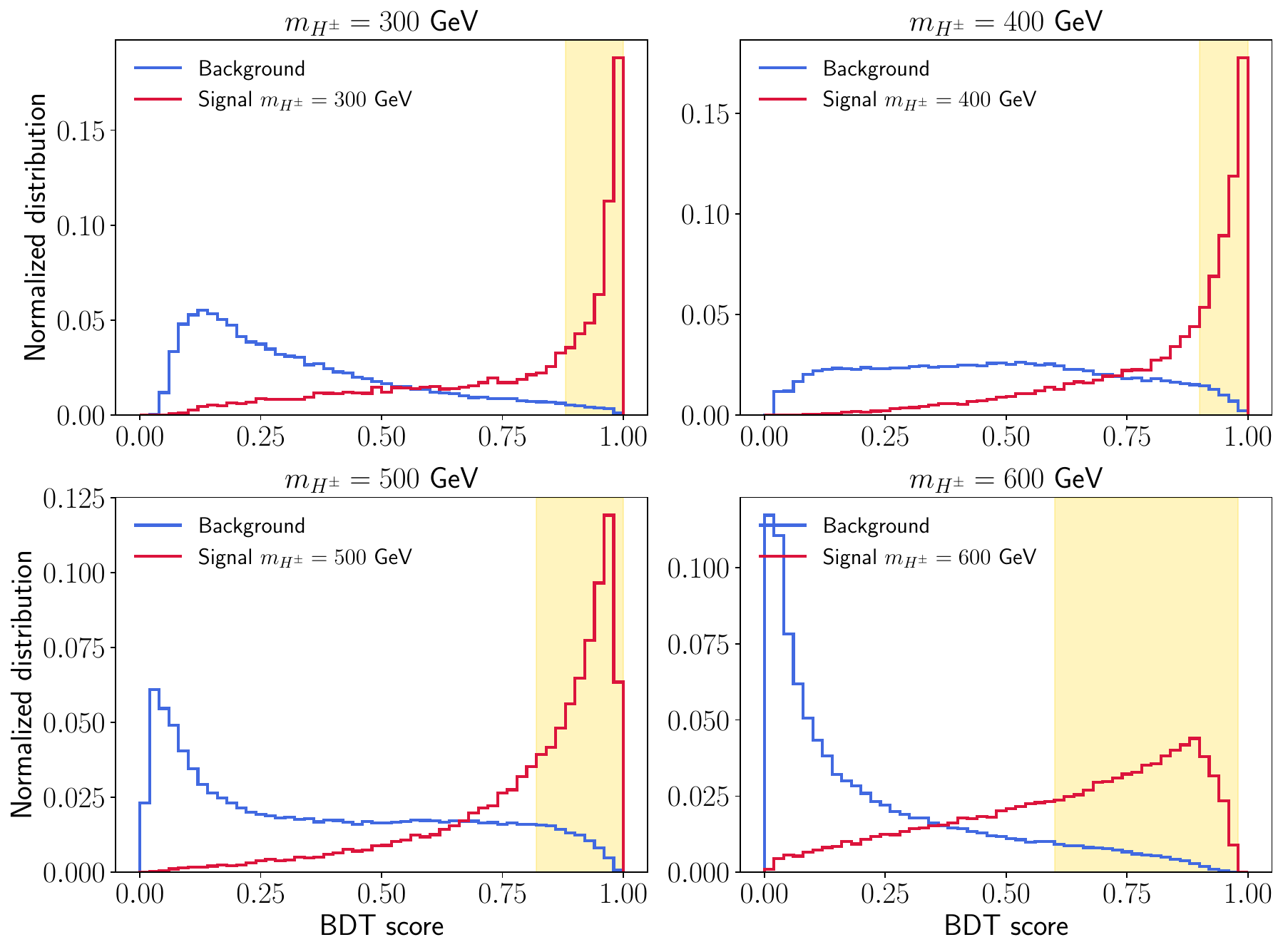}
		\caption{Normalised BDT score distributions for the total background (blue) and the charged Higgs signal (red) at $\tan\beta=5$ for the Type-I scenario. The panels show four representative masses: $m_{H^\pm} = 300, 400, 500,$ and $600$~GeV. The gold shaded region indicates the optimal BDT window that maximises the Asimov significance.}
		\label{fig:bdt_distributions}
	\end{figure}

	The BDT score distributions for the Type-I scenario at $\tan\beta=5$ are shown in Figure~\ref{fig:bdt_distributions}. The signal events populate the high-score region, while the background is concentrated at low scores. A one-dimensional scan over the BDT score is performed to find the optimal window that maximises the profiled Asimov significance $Z$. The chosen windows are highlighted in gold. For higher masses (e.g., $m_{H^\pm}=600$~GeV), the separation power decreases, which is reflected in the broader optimal window.
	
\begin{figure}[htbp]
		\centering
		\includegraphics[width=0.48\textwidth]{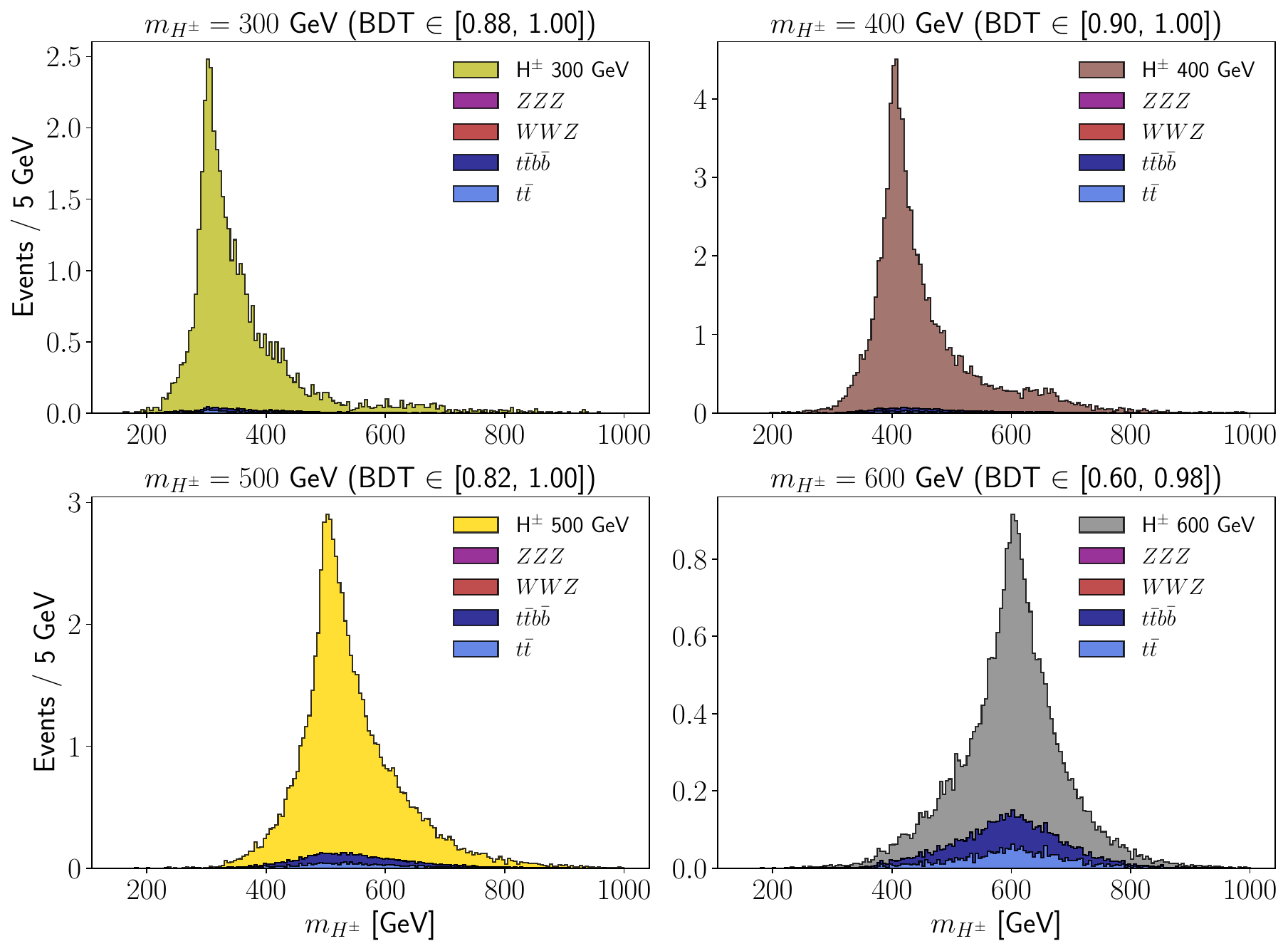}
		\caption{Reconstructed charged Higgs mass $m_{H^\pm}$ after applying the optimal BDT cut for the Type-I scenario at $\tan\beta=5$. The panels correspond to $m_{H^\pm} = 300, 400, 500,$ and $600$~GeV. The signal (coloured) is stacked on top of the remaining background components. The BDT window used for each panel is indicated in the title.}
		\label{fig:mch_after_bdt}
	\end{figure}

	After applying the optimal BDT cut, the resulting $m_{H^\pm}$ distributions are shown in Figure~\ref{fig:mch_after_bdt}. The signal appears as a clear peak at the true charged Higgs mass, while the background is significantly reduced. The residual background is dominated by \ttbar ~and \ttbb ~events that mimic the signal kinematics. 
	
\begin{figure}[htbp]
		\centering
		\includegraphics[width=0.35\textwidth]{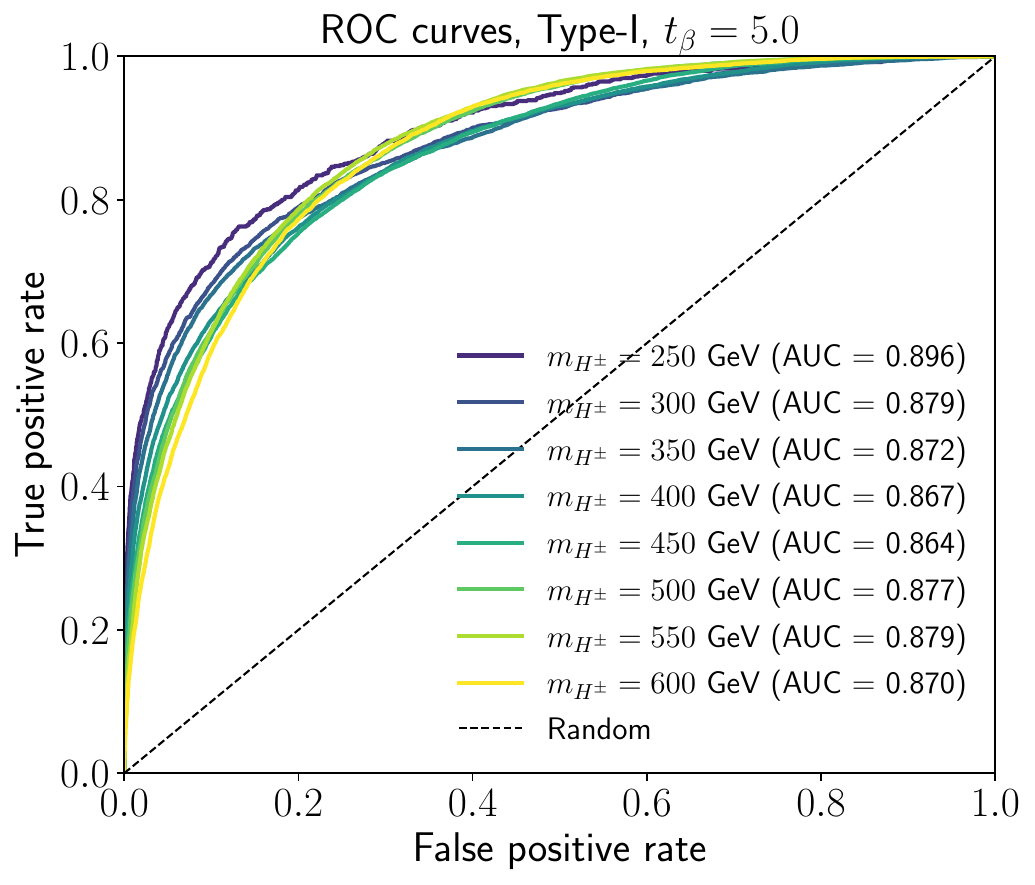}
		\caption{Receiver operating characteristic (ROC) curves for the BDT classifier in the Type-I scenario at $\tan\beta=5$. Each curve corresponds to a different charged Higgs mass hypothesis. The dashed diagonal line represents a random classifier. The AUC values are quoted in the legend.}
		\label{fig:roc}
	\end{figure}

	The performance of the BDT classifier is quantified by the ROC curves shown in Figure~\ref{fig:roc}. The area under the curve (AUC) ranges from $0.864$ to $0.896$ across the mass range, indicating good discrimination power. The AUC is highest for $m_{H^\pm}=250$~GeV and decreases slightly for higher masses, as expected from the reduced production cross-section and the softening of the signal kinematics.
	
	A direct comparison between the cut-based and BDT-based analyses is presented in Table~\ref{tab:z_comparison} for the Type-I scenario. The BDT approach yields a significantly higher significance across the entire mass range. For example, at $m_{H^\pm}=400$~GeV, the cut-based analysis achieves $Z_{\text{cut}} = 12.22$, while the BDT achieves $Z_{\text{BDT}} = 18.71$. The table also lists the number of signal ($N_S$) and background ($N_B$) events in the optimal window for each method. The BDT consistently selects a purer signal region, with a much smaller $N_B$ for a comparable $N_S$.
	
	\begin{table}[htbp]
		\centering
		\caption{Comparison of the profiled Asimov significance $Z$ between the cut-based and BDT-based analyses for the Type-I scenario at $\tan\beta=5$. The columns show the charged Higgs mass hypothesis, the significance and event yields ($N_S$, $N_B$) for the cut-based analysis, and the corresponding values for the BDT analysis.}
		\label{tab:z_comparison}
		\begin{tabular}{c|ccc|ccc}
			\hline\hline
			$m_{H^\pm}$ [GeV] & $Z_{\text{cut}}$ & $N_S^{\text{cut}}$ & $N_B^{\text{cut}}$ & $Z_{\text{BDT}}$ & $N_S^{\text{BDT}}$ & $N_B^{\text{BDT}}$ \\
			\hline
			200 & 0.37 & 0.50 & 1.64 & 0.31 & 2.90 & 55.55 \\
			250 & 3.51 & 13.05 & 8.64 & 7.00 & 13.68 & 1.03 \\
			300 & 8.27 & 47.49 & 14.29 & 14.08 & 40.70 & 1.26 \\
			350 & 11.56 & 80.00 & 16.81 & 18.34 & 69.49 & 1.95 \\
			400 & 12.22 & 96.49 & 21.04 & 18.71 & 77.97 & 2.58 \\
			450 & 11.51 & 107.81 & 30.17 & 16.84 & 78.80 & 4.39 \\
			500 & 9.78 & 83.21 & 27.88 & 14.21 & 65.56 & 5.30 \\
			550 & 6.91 & 48.68 & 23.42 & 10.40 & 41.51 & 5.14 \\
			600 & 3.71 & 24.53 & 25.74 & 5.88 & 20.18 & 5.63 \\
			\hline\hline
		\end{tabular}
	\end{table}
	
\begin{figure}[htbp]
		\centering
		\includegraphics[width=0.48\textwidth]{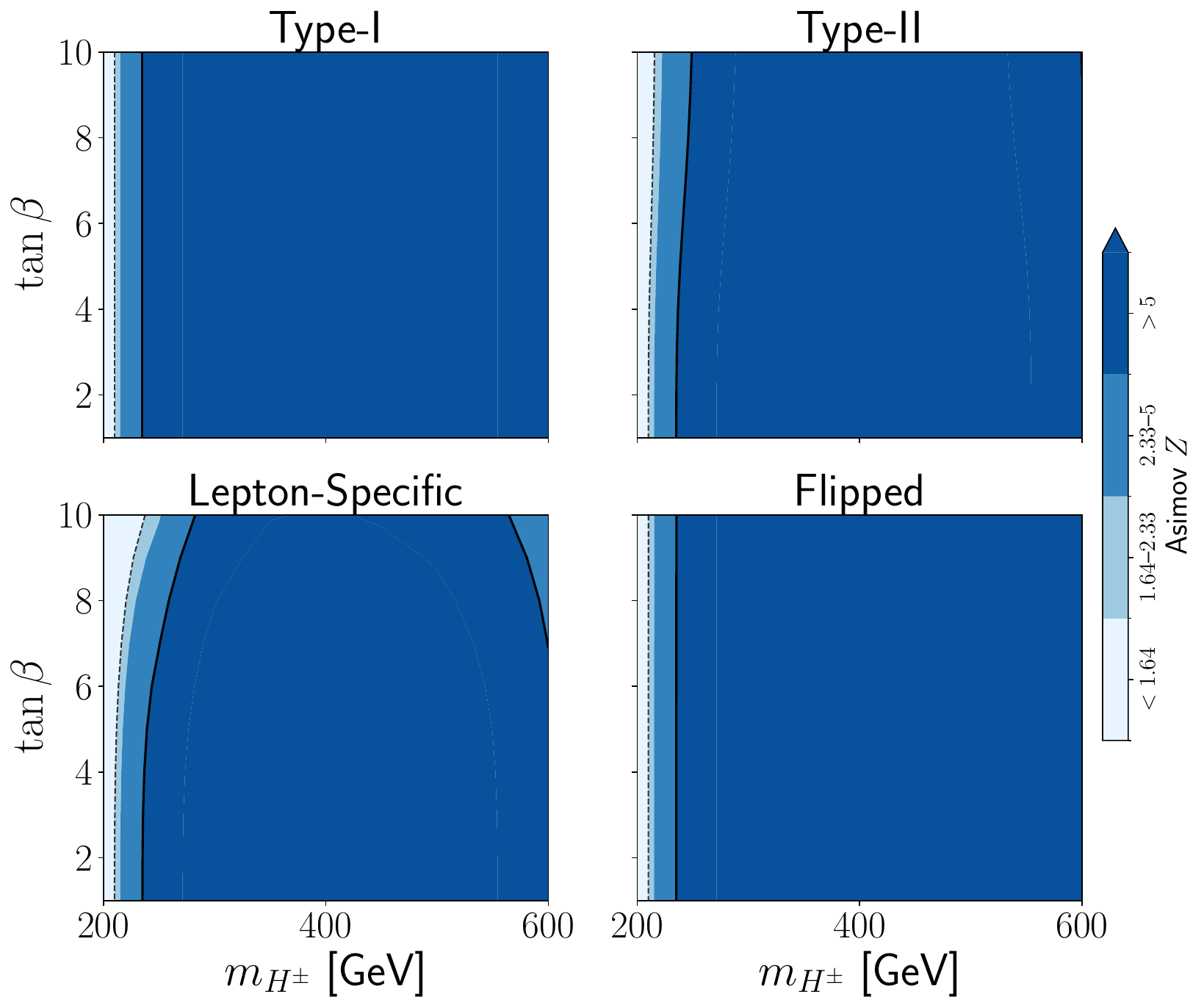}
		\caption{Exclusion contours in the $m_{H^\pm}$--$\tan\beta$ plane for the four 2HDM types (Type-I, Type-II, Lepton-Specific, and Flipped) using the BDT-based analysis. The colour scale shows the profiled Asimov significance $Z$. The solid black contour marks $Z=5$ (discovery), and the dashed contour marks $Z=1.64$ ($95\%$ CL exclusion). The luminosity is $1000$~fb$^{-1}$.}
		\label{fig:exclusion_bdt}
	\end{figure}

	Finally, the exclusion contours in the $m_{H^\pm}$--$\tan\beta$ plane for the four 2HDM types are shown in Figure~\ref{fig:exclusion_bdt}. These contours are obtained by repeating the BDT training and the significance scan across a grid of masses and $\tan\beta$ values. The colour scale indicates the Asimov significance $Z$, with the solid black line marking the $5\sigma$ discovery threshold and the dashed line marking the $95\%$ confidence level exclusion ($Z=1.64$). The BDT analysis extends the reach to more masses and $\tan\beta$ values compared to the cut-based analysis.
	
	\section{Conclusions}
	\label{sec:conclusion}
	
The discovery potential of a heavy charged Higgs boson in the \(H^\pm \to tb\) channel at a 1400~GeV CLIC has been investigated. A cut-based analysis, combining boosted top-tagging with an analytic four-body kinematic fit, shows that a \(5\sigma\) discovery is reachable for \(m_{H^\pm}\) in the range 250--550~GeV at an integrated luminosity of \(1000~\text{fb}^{-1}\). The kinematic fit improves the resolution of the reconstructed charged Higgs mass and provides a robust handle for separating signal from background. It has been explicitly checked that jet-substructure observables (\(\tau_{32}\), \(m_{\rm jet}\)) do not by themselves discriminate the signal from prompt \(t\bar{t}\); the leading discrimination instead comes from the extra \(b\)-jets (the \(\ge 4\) \(b\)-jet requirement) and the reconstructed \(m_{H^\pm}\). To go beyond the cut-based reach, a multivariate analysis based on a boosted decision tree (BDT) was performed. The BDT exploits the full set of kinematic and substructure inputs and substantially improves the separation, extending the \(5\sigma\) discovery reach over almost the entire parameter space considered. This demonstrates that machine-learning techniques are essential for maximizing the physics potential of the \(H^\pm \to tb\) channel at CLIC, and motivates extending the analysis to higher center-of-mass energies and to a full detector-level study.
	\begin{acknowledgments}
		The author thanks the college of science for providing computing facilities.
	\end{acknowledgments}
	
\appendix
\section{Explicit Form of the Correction Factors}
\label{app:correction}

For completeness, the explicit form of the correction factors \(c_i\) is given in this appendix. The four numerators are
\begin{align}
	\mathcal{N}_1 &=
	p_{2z} p_{3y} p_{4x} - p_{2y} p_{3z} p_{4x}
	- p_{2z} p_{3x} p_{4y} \nonumber\\
	&\quad + p_{2x} p_{3z} p_{4y}
	+ p_{2y} p_{3x} p_{4z} - p_{2x} p_{3y} p_{4z},
	\label{eq:N1}\\
	\mathcal{N}_2 &=
	p_{1z} p_{3y} p_{4x} - p_{1y} p_{3z} p_{4x}
	- p_{1z} p_{3x} p_{4y} \nonumber\\
	&\quad + p_{1x} p_{3z} p_{4y}
	+ p_{1y} p_{3x} p_{4z} - p_{1x} p_{3y} p_{4z},
	\label{eq:N2}\\
	\mathcal{N}_3 &=
	p_{1z} p_{2y} p_{4x} - p_{1y} p_{2z} p_{4x}
	- p_{1z} p_{2x} p_{4y} \nonumber\\
	&\quad + p_{1x} p_{2z} p_{4y}
	+ p_{1y} p_{2x} p_{4z} - p_{1x} p_{2y} p_{4z},
	\label{eq:N3}\\
	\mathcal{N}_4 &=
	p_{1z} p_{2y} p_{3x} - p_{1y} p_{2z} p_{3x}
	- p_{1z} p_{2x} p_{3y} \nonumber\\
	&\quad + p_{1x} p_{2z} p_{3y}
	+ p_{1y} p_{2x} p_{3z} - p_{1x} p_{2y} p_{3z},
	\label{eq:N4}
\end{align}
and the common denominator is
\begin{align}
	\mathcal{D} &=
	-E_4 p_{1z} p_{2y} p_{3x}
	+E_4 p_{1y} p_{2z} p_{3x}
	+E_4 p_{1z} p_{2x} p_{3y}
	\nonumber\\
	&\quad -E_4 p_{1x} p_{2z} p_{3y}
	-E_4 p_{1y} p_{2x} p_{3z}
	+E_4 p_{1x} p_{2y} p_{3z}
	\nonumber\\
	&\quad +E_3 p_{1z} p_{2y} p_{4x}
	-E_3 p_{1y} p_{2z} p_{4x}
	-E_2 p_{1z} p_{3y} p_{4x}
	\nonumber\\
	&\quad +E_1 p_{2z} p_{3y} p_{4x}
	+E_2 p_{1y} p_{3z} p_{4x}
	-E_1 p_{2y} p_{3z} p_{4x}
	\nonumber\\
	&\quad -E_3 p_{1z} p_{2x} p_{4y}
	+E_3 p_{1x} p_{2z} p_{4y}
	+E_2 p_{1z} p_{3x} p_{4y}
	\nonumber\\
	&\quad -E_1 p_{2z} p_{3x} p_{4y}
	-E_2 p_{1x} p_{3z} p_{4y}
	+E_1 p_{2x} p_{3z} p_{4y}
	\nonumber\\
	&\quad +E_3 p_{1y} p_{2x} p_{4z}
	-E_3 p_{1x} p_{2y} p_{4z}
	-E_2 p_{1y} p_{3x} p_{4z}
	\nonumber\\
	&\quad +E_1 p_{2y} p_{3x} p_{4z}
	+E_2 p_{1x} p_{3y} p_{4z}
	-E_1 p_{2x} p_{3y} p_{4z}.
	\label{eq:D}
\end{align}
The correction factors are then
\begin{equation}
	c_1 = \frac{s\, \mathcal{N}_1}{\mathcal{D}}, \qquad
	c_2 = -\frac{s\, \mathcal{N}_2}{\mathcal{D}}, \qquad
	c_3 = \frac{s\, \mathcal{N}_3}{\mathcal{D}}, \qquad
	c_4 = \frac{s\, \mathcal{N}_4}{\mathcal{D}},
	\label{eq:c-explicit}
\end{equation}
which is the explicit form of Eq.~\eqref{eq:correction-factors} in the main text. The sign convention is fixed by Cramer's rule with the ordering \(p_1=b_1\), \(p_2=b_2\), \(p_3=t_1\), \(p_4=t_2\), and is the one used in the analysis code.

%
	
\end{document}